\documentclass[10pt,a4paper]{article}
\usepackage{amsfonts,amssymb,amsmath,epsfig,theorem}

\numberwithin{equation}{section}
\newcommand{\R}{{\bf R}}

\newcommand{\Ord}[1]{\ensuremath{\mathcal{O}\!\left(#1 \right)} }
\newcommand{\ord}[1]{\ensuremath{{\text \cal o}\!\left(#1 \right)} }

\theoremstyle{break}

\theoremstyle{break}

\theoremstyle{break}

\theoremstyle{break}

\theoremstyle{break}

\title
{Diffusion  and interfaces  in pattern formation}

\author{Ovidiu Radulescu$^1$, Sergei Vakulenko$^2$,
\\
\\
$^1$IRMAR, UMR CNRS 6625, Universit\'e de Rennes 1, France \\
$^2$Institute for Mech. Engineering Problems Sanct Petersburg,
Russia
}

\begin{document}

\maketitle

\centerline{\bf Abstract}

We discuss several qualitative properties of the solutions of
reaction-diffusion systems and equations of the form $u_t =
\epsilon^2 D \Delta u + f(u,x,\epsilon t)$, that are used in
modeling pattern formation. We analyze the diffusion neutral and the
diffusion dependent situations that, in the time autonomous case,
are distinguished  by considering the attractors of the shorted
equation $u_t = f(u,x)$. We discuss the consequences of being in one
or in the other of the two situations and present examples from
developmental biology and from fluid mechanics.

\section{Introduction}


Mathematical  models in developmental biology \cite{Wolpert,Mjol,
Rein1} and in the theory of phase transitions \cite{Allen, Pom,
RadOlmLu}, lead to reaction-diffusion equations with a spatial
inhomogeneous reaction term of the form $u_t = D \Delta u +
f(u,x)$. There are two different fundamental concepts on the role
of diffusion in development (patterning) and, respectively, two
different approaches.

The first, more mathematical, approach is pioneered by the work of
Turing \cite{Turing} and developed by many authors
(see, for example, \cite{Meinhardt, Murray}). For Turing's
patterning mechanisms \cite{Turing, Meinhardt} a spatial
dependence of the reaction term $f=f(u,x)$ is not necessary.
Conditions on  diffusion coefficients and on
 reaction terms are needed in order to have
spontaneous translation symmetry breaking by developing periodic
patterns from a homogeneous state \cite{Hun}.

The second approach is inspired from the ideas of the
19$^{\text{th}}$ century biologist Hans Driesch who pointed out that
the fate of a cell can be determined by its position inside the
embryo. Later, it has been shown that differentiation is connected
with reading an inhomogeneous level (gradient) of maternal proteins
(morphogens). Wolpert \cite{Wolpert,wolpert-waddington} coined the
term "positional information" and gave the theoretical basis of
gradient models. In a short discussion of morphogenesis, Ren\'e Thom
\cite{thom-waddington} proposed that spatially inhomogeneous
reaction-diffusion equations can model formation of patterns
containing interfaces. He pointed out that interfaces are placed
according to the Maxwell rule \cite{thom-waddington,Pom,RadOlmLu}.
Nevertheless, biologists seem to reject the role of diffusion in
Wolpert's mechanism.
This belief is backed up by the fact that without diffusion monotone
morphogen profiles can induce complex layered patterns consisting of
many narrow interfaces \cite{Wolpert}.
In fact, one can show that patterns of arbitrary complexity can be
generated by the diffusionless Wolpert mechanism
\cite{VakGrCR,Gen8}.

Turing and Wolpert type mechanisms are not restricted to
developmental biology. They apply to other pattern formation
phenomena. Thus, in material science and fluid mechanics Cahn's
spinodal instabilities are similar to Turing instabilities from
biology and can be modeled by space homogeneous partial differential
equations \cite{Cahn-Hilliard}. Other examples of Wolpert-like
mechanisms can be found in fluid mechanics. Flow-induced phase
transitions can trigger the formation of bands separated by narrow
interfaces \cite{OlmRadLu,RadOlmLu,RadOlm,Rad}.



These examples suggest that interfaces play an important role in
patterning. An interface can be defined as a region of strong
inhomogeneity. Fick's law  implies that diffusion is important at
interfaces, but it says nothing on the role of diffusion for the
control of the interface position and dynamics. In particular, how
do the patterns behave in the limit of vanishing diffusion?

Two distinct situations are considered here.
 In the first situation,
patterning follows essentially from the nonlinearity of the reaction
term and does not suffer qualitative changes  in the limit of
vanishing diffusion. The corresponding patterning mechanism will be
called diffusion neutral. In the second situation, patterning
depends essentially on diffusion and the patterning mechanism will
be called diffusion dependent. These two types of patterning are
related to  two different situations and, respectively,
 two types of interfaces.

The first situation can appear, for example, when dynamics of  a
"shorted system" (where the diffusion is removed) has, as a global
attractor, for any values of space variables $x$, a stable rest
point (depending on $x$). In this case,  theorems on a connection
between shorted system and the original reaction diffusion system
are established. The  interface existence results from the steep
spatial dependence of the reaction term on $x$.

The second situation arises, when the two local attractors coexist.
In this case, patterning cannot be diffusion neutral and thus it is
incorrect to remove the diffusion in the equations. Here an
interface appears as a localized jump in space from one attractor to
another.

There exists a rather well developed mathematical theory for the
second situation in the one-component case (see \cite{Carr, Fusco,
Fife, MOLV} among many others), if local attractors are rest points.
In this case the  interfaces are well localized traveling wave
solutions of singularly perturbed
 reaction-diffusion equations. Stationary patterns are formed by
interfaces corresponding to  zero velocity traveling waves. Such
stationary interfaces occur in many systems and they are well
represented in biology (see \cite{Murray}) and  in the phase
transition theory \cite{OlmRadLu, RadOlmLu, RadOlm, Rad, Pom}. Their
existence is well known for reaction-diffusion equations with a
typical bistable nonlinearities and for some reaction-diffusion
systems \cite{VVA1,VVA2,VVA3}. Unfortunately, up to now there is no
general theory for
 systems of reaction-diffusion equations, except for rather
restrictive situations (gradient systems and monotone systems,
\cite{VVA1,VVA2,VVA3}).

Notice that interfaces can interact one with another.
Under some conditions, one can reduce the pattern dynamics (that is
a dynamics
 in an infinite dimensional space) to the finite dimensional dynamics of
a set of interacting interfaces \cite{MOLV,Carr, Fusco}.
If this reduction is possible, the result is a
 simplification of pattern time
evolution. Instead of direct numerical
simulation of patterning, one can study time evolution  of
interfaces that can be described by ordinary differential equations
(instead of partial differential equations). In many cases,
these ordinary equations can be resolved analytically.
In this paper we also give  new results on interface motion and
apply these results to genetic circuits and to fluid mechanics.

\section{Statement of the problem}
We  study initial value boundary problems for  the following
reaction-diffusion systems:
\begin{equation}
u_t = \epsilon^2 D \nabla^2 u + f(u,x,\epsilon t), \label{eq2.1}
\end{equation}
where $u=u(x, t) \in{\bf R}^n$ is an unknown vector function,
$x\in {\Omega} \subset {\bf R}^q$, where $\Omega $ is a compact
domain with a regular boundary $\partial \Omega$, $ t \ge 0$. In
equation \eqref{eq2.1},  $D$ is a diagonal matrix with positive entries,
i.e., $D=diag\{d_1, d_2, ..., d_n\}$, $\epsilon$ is a parameter for
diffusion intensity, $f$ is a reaction term depending, in general,
on $x$ and also slowly on time.
 Initial data has
the following form
\begin{equation}
u(x,0)=u_0(x). \label{eq.2}
\end{equation}

In applications, boundary conditions
are usually of the Neumann type. This means  that there is no flux
across the boundary:
\begin{equation}
\nabla u(x) \cdot n(x)  = 0, \quad x \in  \partial {\Omega},
\label{eq2.3}
\end{equation}
where $n(x)$ is an unit normal vector to the boundary at $x$. This
work is focused on internal layers, and to avoid complicated problem
connected with  boundary layers, we sometimes consider a simplified
problem, when $\Omega$ is a box in ${\bf R}^q$, with periodical
boundary conditions in $x$.

In this paper we  refer to  the gene circuit model proposed to
describe early stages of Drosophila (fruit-fly) morphogenesis
\cite{Mjol, Rein1}. In this model the reaction term takes a special
form (that reminds the Hopfield neural network model):
\begin{equation}
f_i =\sigma_{\alpha} (\sum_{j=1}^n K_{ij} u_j  + \sum_{k=1}^p J_{ik}
m_k(x) -h_i) - \lambda_i u_i,
\label{eq2.4}
\end{equation}
where $u_j$ are zygotic gene concentrations, $K$ is a matrix
describing pair gene interaction between zygotic genes, $J$ is a
matrix describing pair interaction between zygotic genes and
maternal genes,
 $h_i$ are thresholds, $m_i$ are functions of $x$
which define maternal gene concentrations (morphogen gradients).
Here $\sigma_{\alpha}(h)=\sigma(\alpha h)$, $\sigma$  is a  monotone
and smooth (at least twice differentiable) "sigmoidal" function such
that
\begin{equation}
\sigma(-\infty)=0, \quad \sigma(+\infty)=1.
\label {eq2.5}
\end{equation}
 The function $\sigma_{\alpha}$ becomes a step-like function as
its sharpness $\alpha$ tends to $\infty$. Typical examples can be given
by
\begin{equation}
     \sigma(h)=\frac{1}{1 + \exp(-h)}, \quad \sigma(h)
     =   \frac{1}{2} \left( \frac{h}{\sqrt{1+h^2}} + 1 \right).
\label {eq2.6}
\end{equation}

Slow dependence of $K,J,h,\lambda$ on time can be also considered.


Our second example is a generalization of the Allen-Cahn model
 of phase transitions. The spatial
homogeneous version of this model has been discussed in connection
to equilibrium first order phase transitions \cite{Allen},
population dynamics \cite{Gurtin}, metastability phenomena
\cite{Carr,Pinto}. It has been used  as a toy model to describe
shear banding of complex fluids \cite{RadOlmLu}. The spatial
inhomogeneous model can be described by the following
reaction-diffusion equation with Neumann no flux boundary conditions
on some compact set $\Omega \subset \R^q$:
\begin{equation}
 u_t = \epsilon^2 u_{xx} - A^2(x,\tau) [u - u_0(x,\tau)][ u - u_2(x,\tau)][
 u - u_1(x,\tau)],
\label {eq2.7}
\end{equation}
where $\tau = \epsilon t$, $u_0(x,\tau) < u_2(x,\tau) <
u_1(x,\tau)$ and $u_i(x,\tau), A(x,\tau)>0$ are smooth (at least
$C^2$), real functions.


\subsection{Main ideas}


We are interested in the influence of diffusion on patterning. We
shall call a pattern formation mechanism "diffusion neutral", if the
solution $u^{\epsilon}(x,t)$ of system (2.1) converges uniformly
to a pattern $v(x,t)$ that can be found in absence of diffusion, as
the diffusion coefficients converge to zero. In this case the
diffusion term is a regular perturbation. The pattern formation is
"diffusion dependent", if this does not hold.

More precisely, let us formulate the following definition.

Let $u^{\epsilon}(x,t)$ denote the solution of
reaction-diffusion system (2.1) with Neummann conditions (2.2) or
with periodical boundary conditions in $x$ on $\Omega = [0, 1]^q$
and initial data $u^{\epsilon}(x,0)=u_0(x)$.


Let us introduce the following system of ordinary differential
equations (we shall refer to it as the "shorted" system):
\begin{equation}
v_t = f(v,x,  t),  \quad v(x,0)=u_0(x).
\label{eq2.8}
\end{equation}

{\bf Definition}. {\em Patterning defined by  problem (2.1), (2.2)
with initial data $u_0(x)$ is diffusion neutral, if the
estimate
$$
    |u^{\epsilon}(x,t) - v(x,t)|  < r_{\epsilon}
\label {eq2.9}
$$
  holds for any $x \in \Omega, t > 0$, where $r_{\epsilon} \to 0$ as
$\epsilon \to 0$. The number $r_{\epsilon}$  is independent of $x,t$ but
can depend on initial data.
Otherwise,  patterning is diffusion dependent}.

Notice that patterning can be diffusion neutral or diffusion
dependent depending on initial data $u_0(x)$.

In biological axial morphogenesis and fluid mechanics shear induced
phase transitions, typical patterns are segmented, with narrow
interfaces between segments \cite{Wolpert,RadOlmLu}. We show that
the "diffusion neutral" and "diffusion dependent" situations
correspond to two different types of interfaces. To illustrate the
main ideas, let us consider the one-dimensional and one-component
case. ($q=1, n=1$).

{\bf Stationary patterns}

An important situation in pattern formation is the case, where
system (2.1) has steady state solutions. In biology, steady states
are always approximations, valid within a certain time scale. In
order to study stationary patterns, let us consider the autonomous
case of system (2.1), where $f$ does not depend on $t$.

\begin{figure}

\begin{minipage}{10cm}

\centerline{
\includegraphics[width=5cm]{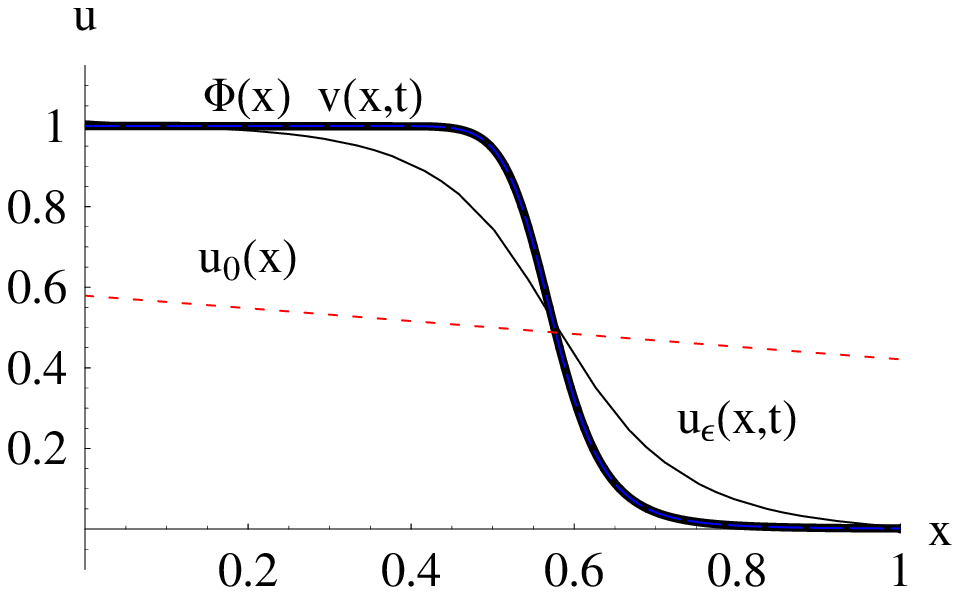}
\includegraphics[width=5cm]{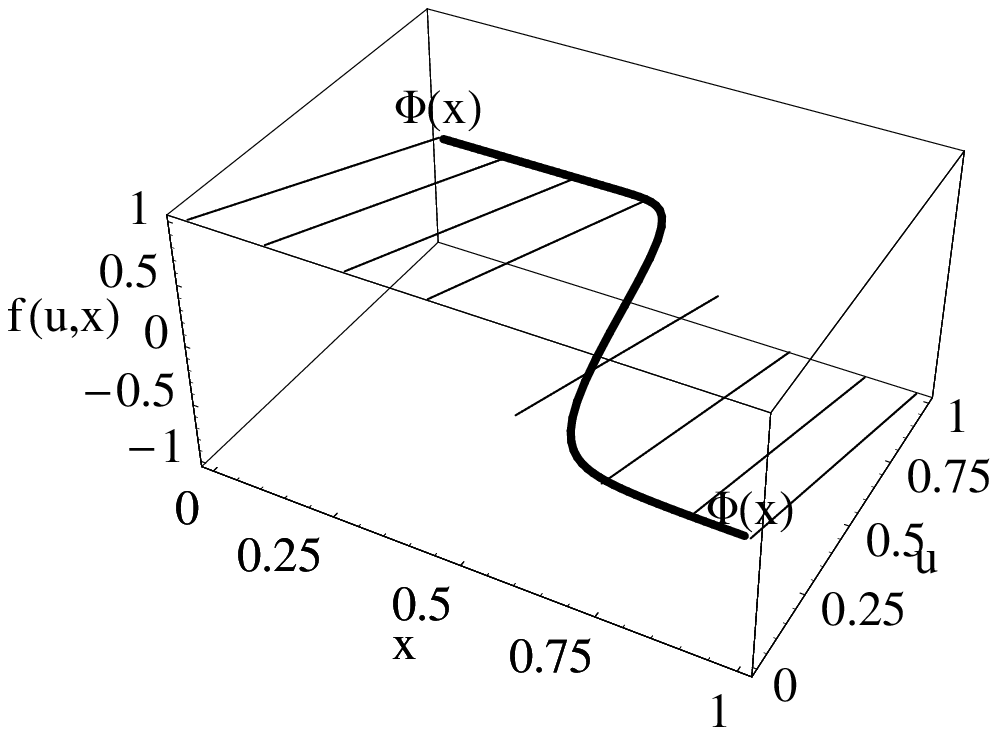}}

\centerline{ a)  }

\centerline{
\includegraphics[width=5cm]{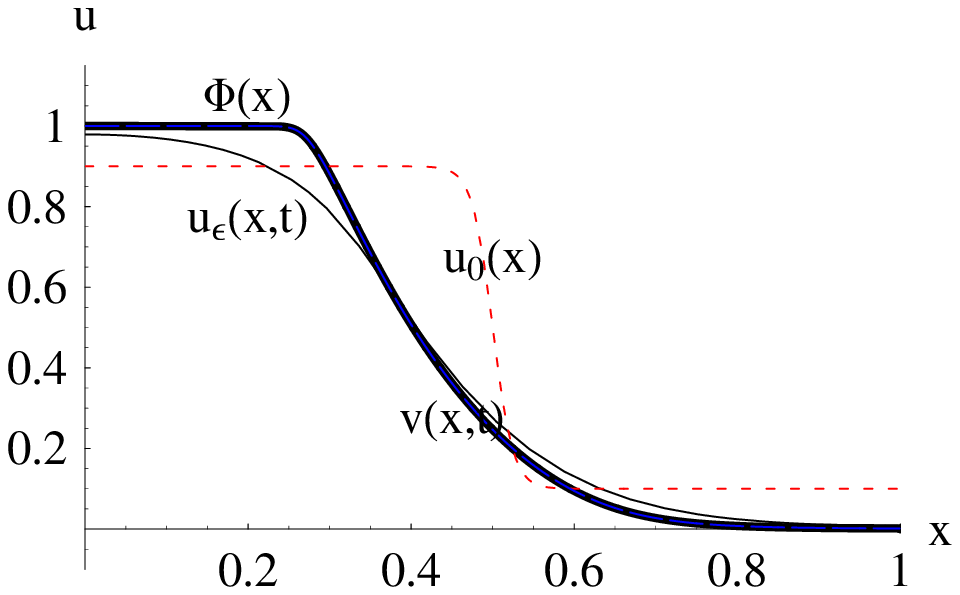}
\includegraphics[width=5cm]{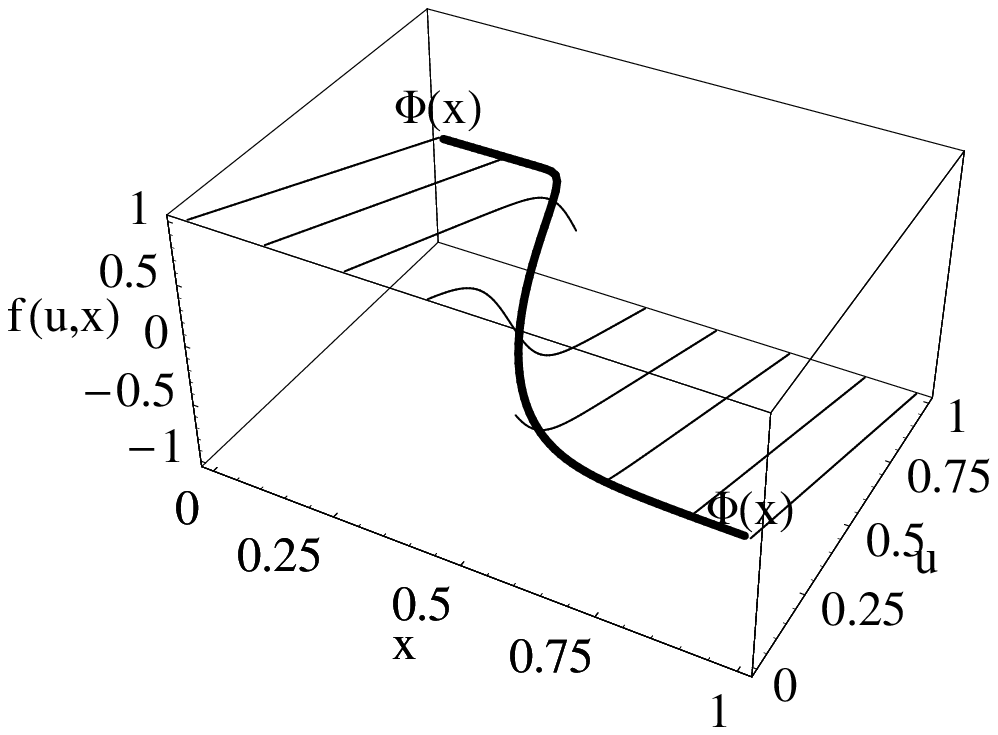}}

\centerline{ b)  }

\centerline{
\includegraphics[width=5cm]{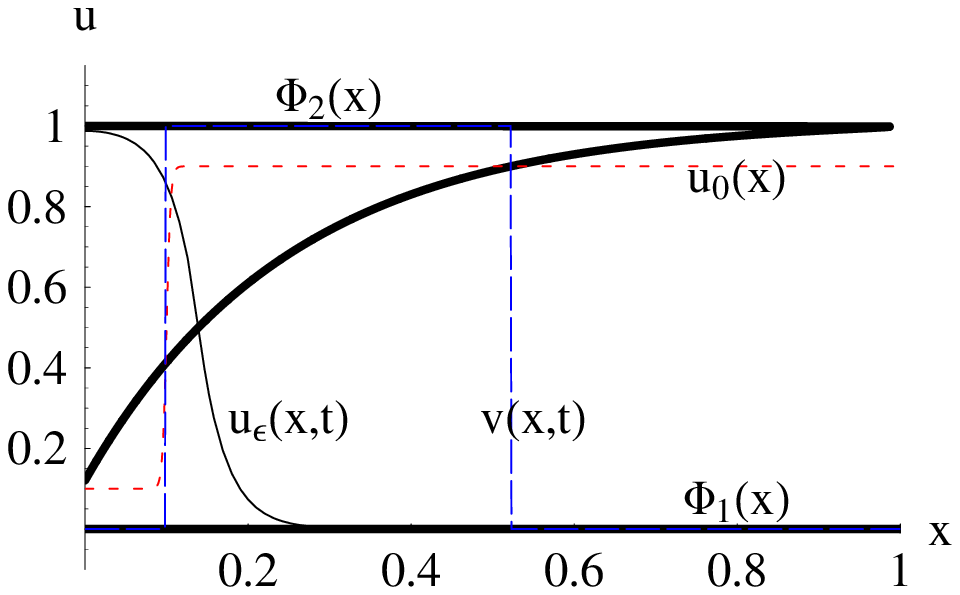}
\includegraphics[width=5cm]{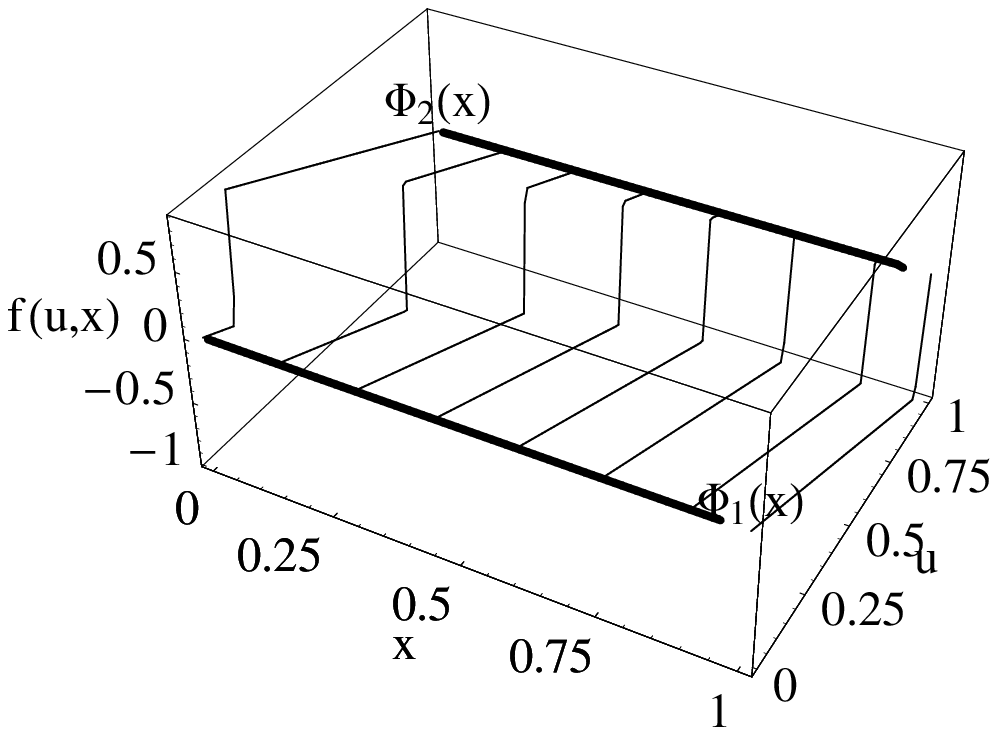}}

\centerline{  c) }

\centerline{
\includegraphics[width=5cm]{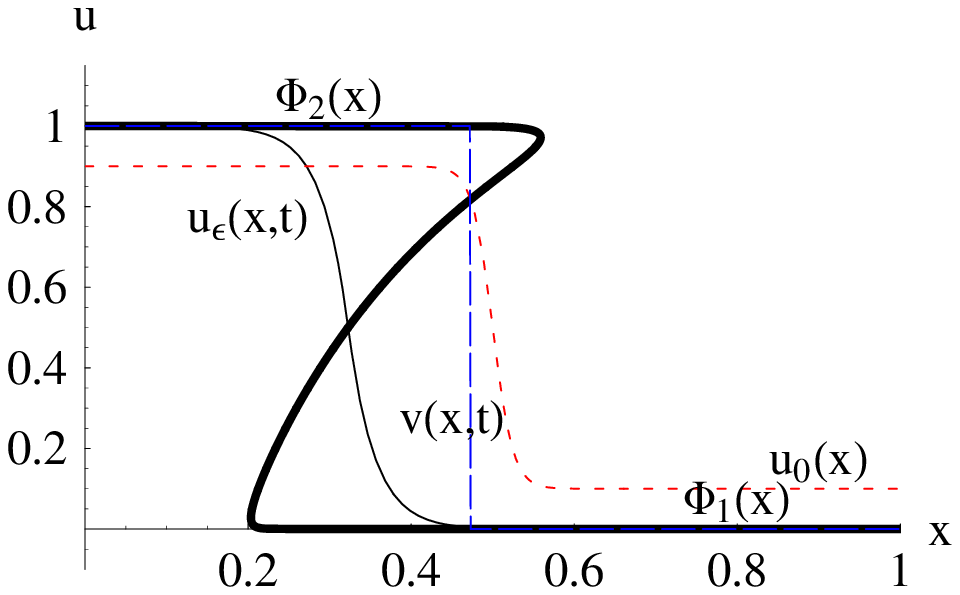}
\includegraphics[width=5cm]{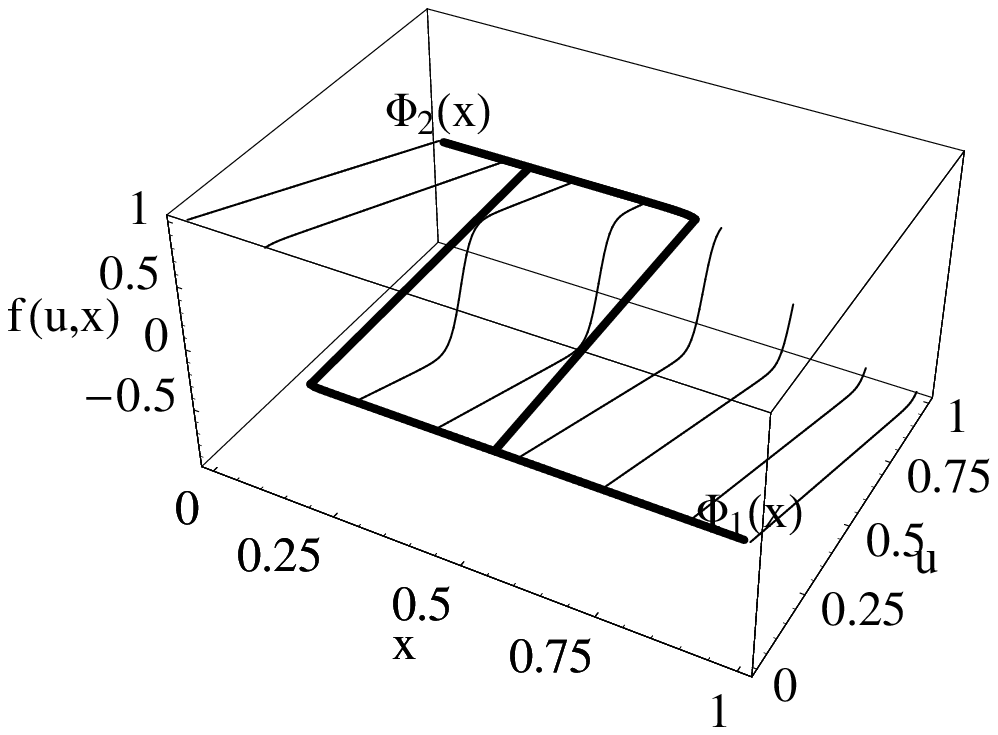}}

\centerline{ d) }

\caption{Gene circuit model for
$n=1,p=1,q=1,\sigma_{\alpha}(x)=1/(1+exp(-\alpha x))$,
$m(x)=exp(-x)$. Diffusion neutral patterning: a) $\alpha = 80.0$,
$K_{11} = 0$, $J_{11} = 1.0$, $h_1 = 0.1$, $\epsilon^2=0.1$. b)
$\alpha = 80.0$, $K_{11} = -0.2$, $J_{11} = 1.0$, $h_1 = 0.1$,
$\epsilon^2=0.1$. Diffusion dependent patterning: c) $\alpha =
400.0$, $K_{11} = 1.7$, $J_{11} = 1.5$, $h_1 = 1.71$,
$\epsilon^2=0.001$. d)  $\alpha = 80.0$, $K_{11} = 0.45$, $J_{11} =
1.0$, $h_1 = 0.5$, $\epsilon^2=0.001$. The solution
$u_\epsilon(x,t)$ of the reaction-diffusion equation and the
solution $v(x,t)$ of the shorted equation are represented after a
large time when the pattern is practically at the attractors. In the
diffusion dependent case, depending on initial data $u_0$ the
solution $v(x,t)$ of the shorted equation can have many interfaces
(two in c)). }
\end{minipage}
\label{fig1}
\end{figure}

There occur  two essentially different cases that are illustrated in
Fig.\ref{fig1}:

{\em Case I}. For any  $x$, the shorted
equation has an unique point attractor $v=\phi(x)$,
where $\phi$ is a solution of the equation
$f(\phi(x),x)=0$. This attractor, for each $x$, attracts globally all
trajectories of the shorted system.


For small $\epsilon$, the solution $u^{\epsilon}(x,t)$ is close to
the solution $v(x,t)$  of the shorted equation. This function $v$
tends to $\phi(x)$ for large times. In this case a narrow interface
can occur only if the function $f(\phi, x)$ is, in a sense, "sharp"
in $x$. We shall refer to such a region as  type 1 interface or
"transition layer", because the variations of one or several
components of the function $\phi(x)$ are strong across it.
%
It is the case, for example, if, in the Eq. (2.4), the parameter
 $\alpha$ is large. For $n=p=q=1$, $J_{11}m_1(x)=k x$, $K_{11}=0$,
the solution $\phi$ has the
following form
\begin{equation}
\phi(x) = \frac{A }{1 +  exp(-\alpha (k x -h_1))}, \quad A >0.
\end{equation}
This solution describes an interface at $x=h_1/k$ whose width is
$1/(k\alpha)$.

{\em Case II}. There exist intervals in $x$, where the shorted
equation $f(u,x)=0$ has several different attractors, for example
$\phi_1(x), \phi_2(x),...,\phi_k(x)$. The interfaces correspond to a
localized jump from an attractor to another one, the functions
$\phi_k(x)$ having otherwise slow variation. In this case interfaces
have widths of order $\epsilon$, and are diffusion controlled. We
shall refer to these as  type 2 interfaces or "diffusion layers".

In the case II, diffusion neutral patterning is possible if initial
data $u_0(x)$ lies in an attraction basin of only one attractor, for
instance $\phi_1(x)$. If initial data lies in the attraction basins
of different attractors, the pattern contains internal layers and
patterning is diffusion dependent. For small $\epsilon$, the
solution $u^{\epsilon}(x,t)$ is close to the solution of the shorted
equation only for times $\Ord{-\log(\epsilon)}$. The dynamics is
diffusion dependent because, in the absence of diffusion the
derivatives of the solution of the shorted equation increase
unboundedly at some $x$ \cite{MOLV}. We can identify these $x$ with
the initial positions of the diffusion layers.

In the presence of diffusion, interfaces move toward diffusion
controlled equilibrium positions that are generally different from
the positions in the absence of diffusion \cite{FifeHsi,RadOlmLu,
RadOlm}. Let us notice that the steady state solutions of the
shorted equation can have an arbitrary number of interfaces in
arbitrary positions determined by the sequence of attractor basins
to which the initial data belongs to. With diffusion, there are
restrictions on the number and on the positions of the interfaces
\cite{RadOlm}.

\subsection{Main results}
Our main result for diffusion neutral patterning  are Theorems 3.1
and 3.2 that will be formulated in the next section. Let us
summarize this result here. Let us suppose that in the autonomous
case $f=f(u,x)$ shorted system (2.8) has only the point
attractors (equilibria) $\phi_k(x)$, with open attraction basins
$B_k(x)$ for all $x$. The case of bifurcations and turning points,
where several curves $\phi_k(x)$ touch one another
\cite{thom-waddington} needs special treatment and will not be
considered in this paper. Furthermore, let us assume that the
initial data $u_0(x)$ is contained in $B_1(x)$ for any $x$. Then
the solution $u^{\epsilon}(x,t)$ of  system (2.1) with
periodic boundary conditions stays, for all times $t$, in a small
neighborhood of the solution $v(x,t)$ of the shorted system.
Naturally, $v(x,t) \to \phi_1(x)$ and, therefore,
\begin{equation}
u^{\epsilon}(x,t) = \phi_1(x) + o(\epsilon^s), \quad
\epsilon \to 0, s>0
\label {eq2.11}
\end{equation}
for sufficiently large $t > t_{\epsilon}$.

Let us consider now the diffusion dependent patterning.
Recently it has been shown that reaction-diffusion systems can
generate any structurally stable dynamics. This result holds even
for the two component case, if $f$ is allowed to depend on $x$
\cite{Vak2,Vak5}. Arbitrarily complex dynamics can be reproduced by
genetic circuit systems defined by systems (2.1) with nonlinearities
 (2.4)
\cite{Fun, Vak6}. Furthermore these systems are capable to produce
any space-time patterns depending on the parameters $n, K, h$
\cite{Gen8, VakGrCR}.


If, for some $x$, the shorted system possesses stable limit cycles
or chaotic  attractors, we can prove only a weak variant of
Theorem 3.1, where $u^{\epsilon}$ stays close to $v$ only on a large
but bounded time
interval $O(-\log \epsilon)$. It is difficult to say something
beyond this interval. If the attractor of the shorted system is a
stable limit cycle, rigorous results are unknown, however, a
formal asymptotic solution can be obtained by the Kuramoto method
\cite{Kuramoto}. This solution shows
that, for generic initial data, patterning is diffusion dependent.
Nothing is known on the structure of the solution
$u^{\epsilon}$ for chaotic attractors. However, taking into account
that such attractors contain many periodic trajectories, one can suppose
that in this case patterning is diffusion depending as well.

An interesting
situation arises if initial data $u_0$ can lie, for different $x$,
in attraction basins of different attractors $\phi_k(x)$.
For systems we prove here a general theorem which shows that
in this case  patterning is diffusion dependent, but it is difficult
to describe patterning analytically.

An analytical approach is possible for reaction-diffusion equations.
The general case of reaction-diffusion equations with a spatial
inhomogeneity has been considered by \cite{Fife,FifeHsi}. These
works gave a description of stationary internal layers and
convergence to these layers. Our Theorem 5.2 extends some results of
\cite{Fife}. For the one-component, one-dimensional patterning
problem  we prove existence of traveling waves in the bistable case
and we reduce the patterning dynamics to a differential equation for
the position of the inner layer. We also include a slow time
dependence of the reaction term on timescales slower than the
diffusion time.


\section{Diffusion neutral patterning}

First, let us we formulate some assumptions.

Suppose that the reaction term $f$ is a $C^2$ regular function
of $x, u$ and the initial data $u_0$
have at
least $C^2$-regularity on the domain $\Omega$. Let us consider the case
when $\Omega$ is a compact box in ${\bf R}^m$ and let us set periodical
boundary conditions along all its edges.

Our next assumption is concerned with the attractors and the
attractor basins of shorted system (2.8). Notice that
(2.8) is a system of ordinary differential equations in
${\bf R}^n$ whose right hand sides and initial data involve  $x$ as
a parameter. Let us assume that for each $x$  dynamics (2.8)
 is dissipative, i.e., there exists a ball $B_R$ of
a radius $R$ centered at $v=0$ and this ball is a globally attracting
set. This is the case, for example, if there holds the following

{\bf Dissipativity condition for shorted system:} {\em For any
$x\in \Omega$ there exists a positive number $R(x)$ such that

\begin{equation}
u \cdot f(u,x)  < 0 \quad (|u|=R(x), \ x \in \Omega)
\label{eq3.1}
\end{equation}}

Sometimes we shall use  the stronger

{\bf Uniform Dissipativity condition:} {\em There exists a positive
number $R_0$ such that for any $x$
\begin{equation}
u \cdot f(u,x)  < 0 \quad (|u|=R_0, \ x \in \Omega)
\label{eq3.2}
\end{equation}}

This assumption guarantees  existence and uniqueness of solutions of
the problem (2.1),(2.2). It is easy to show that this condition
holds for genetic  circuits.

Let us recall the definition of local attractor and of their
attraction basin for dissipative systems of ordinary differential
equations. Let us assume that the dissipativity condition holds.
For a fixed $x$, the set ${\cal A}_x$ is  a local attractor for
the shorted system, if this set has an  attraction basin $B({\cal
A}_x)$ containing an open neighborhood $V_x$ of ${\cal A}_x$. The
attraction basin $B({\cal A}_x)$ consists of all $u_0(x)$ such
that the trajectory $u(t,x,u_0(x))$ starting at $u_0(x)$
approaches to ${\cal A}_x$ for large times:
$$
   \text{dist} [ u(t,x, u_0(x)), {\cal A}_x ] \to 0 \quad (t \to  \infty).
$$

Suppose among the local attractors of (\ref{eq2.6}) there exists at
least one isolated branch (parametrized by $x$) of point attractors
${\cal A}_x = \{ \phi(x) \}$. This means that $\phi(x)$ are stable
steady states of the shorted system.

{\bf Attraction basin condition:} {\em Suppose that for each $x$
there is a point attractor ${\cal A}_x=\{\phi(x) \}$ of the system
(2.8). Moreover, assume that the initial data $u_0(x)$ lie, for all
$x$, in the corresponding basin $B({\cal A}_x)$:
\begin{equation}
  u_0(x) \in B({\cal A}_x)=B(\{\phi(x)\}).
\label{eq3.3}
\end{equation}}

Let us formulate a condition for the stability of ${\cal A}_x$. Let
$M(x)$ be the derivative of $f$ at $\phi(x)$:

\begin{equation}
f(\phi + w, x) -f(\phi, x) =M(x) w + h(w,x), \quad |h| < C|w|^2
\label{eq3.4}
\end{equation}
 for small $w$ and some constant $C$.
 Suppose  that  there holds the  following

{\bf Strong linear stability assumption:} {\em For any $x \in
\Omega$ the corresponding matrix
  $M(x)$ at $\phi(x)$ satisfies the following condition:
\begin{equation}
  \sum_{j \ne i} |M_{ij}(x)| +  M_{ii}(x) \leq -b <  0.
 \label{eq3.5}
\end{equation}}

This condition
 ensures that the spectrum of the matrix $M(x)$ lies
in the negative half-plane being separated by a gap from the
imaginary axis. This gap is uniform in $x$ due
to the compactness of $\Omega$. Hence,  $\{\phi(x)\}$ has an open
attraction basin $B_x$ for all $x$.

We can state now

{\bf Theorem 3.1}. {\em Let $f=f(u,x)$ and $u_0$ are sufficiently
smooth as formulated above.

Under the uniform dissipativity, strong linear stability and
attraction basin conditions
 for sufficiently small $\epsilon$, the solution $u^{\epsilon}$
of the time autonomous version of problem  (\ref{eq2.1}), (2.2)
 exists for all
$t > 0$ and stays close to the solution of the shorted system (2.8),
i.e.
\begin{equation}
u^{\epsilon}(x,t) = v(x,t)  + \tilde v_{\epsilon}(x,t),
\label {eq3.6}
\end{equation}
where the correction $\tilde v^{\epsilon}$ satisfies the estimate
\begin{equation}
\label{eq3.7}
|\tilde v^{\epsilon}(x,t)| < c\epsilon^{s}, \quad t \geq 0,
\end{equation}
where $c > 0$ is a constant independent of $\epsilon$, $s > 0$.

Furthermore, for sufficiently large times $t > t_{\epsilon}$ the
following result holds:

\begin{equation}
  u^{\epsilon}(x,t) =  \phi(x) + \tilde u^{\epsilon} (x, t),
\label{eq3.8}
\end{equation}
where $\tilde u^{\epsilon} (x,t)$ satisfies the following estimate
uniformly in $x$:
\begin{equation}
|\tilde u^{\epsilon} (x, t)| < c_1 {\epsilon^2}.
\label{eq3.9}
\end{equation}
}

{\bf Proof}. The proof uses the following Lemma, which is a slightly
modified version of a comparison theorem for systems of
reaction-diffusion equations \cite{Smoller}. To formulate
this Lemma, we introduce two sets depending on two vectors $w^-,
w^+$. We denote
$$
E^i_+(w^+,w^-)   = \{ \xi |
\xi_i = w^+, w^- \leq \xi_j \leq w^+, \forall j \neq i \},
$$
$$E^i_-(w^+,w^-)   = \{ \xi | \xi_i = w^-, w^- \leq \xi_j \leq w^+,
\forall j \neq i \}.
$$
\vspace{0.2cm}


{\bf Lemma 3.1}
{\em
Let $w(x,t) \in {  \Bbb R }^n$ be the solution of problem (2.1), (2.2)
 with zero Neumann or periodic boundary conditions and
initial data $w(x,0)=w_0(x) \in {\Bbb R}^q$.
Let the time dependent functions $w^+(t),w^-(t)$ satisfy
$$
w_t^+ \geq  \max_i \sup_{\xi \in
E^i_+(w^+,w^-)} f_i(\xi,x),
$$
$$w_t^- \leq \min_i \inf_{\xi \in
E^i_-(w^+,w^-)} f_i(\xi,x), \,\text{for any} \, \  x \in \Omega,
$$
where $1
\leq i \leq n.$

Moreover, let us suppose
$$w^-(0) \leq w_i(x,0) \leq w^+(0), \,\text{for any}\,  \ x \in \Omega.$$

Then
$$w^-(t) \leq w_i(x,t) \leq w^+(t), \, \text{for any} \,\, t > 0,
\ x\in \Omega. $$
}

To simplify the proof, we proceed with it in  two parts, I and II.
In the first part we show that $u_\epsilon(x,t)$ stays within
distance $\ord{\epsilon^s}$ from the solution $v(x,t)$ of the
shorted system for times $t < t_\epsilon = -a \log(\epsilon)$. We
use the exponential decay of $v(x,t)$ to $\phi(x)$ in order to show
that at $t= t_\epsilon$, the solution $u_\epsilon(x,t)$ is within distance
$\ord{\epsilon^s}$ from $\phi(x)$. In the second part we apply
again the lemma and find that $u_\epsilon(x,t)$ remains within
distance $\ord{\epsilon^s}$ from $v(x,t)$ for all $t \geq t_\epsilon$.

{\bf Part I}. Let $w(x,t)=u^{\epsilon}(x,t) - v(x,t)$,
where $v(x,t)$ is the solution of the shorted system with the same
initial data as $u_\epsilon(x,t)$, $v(x,0)=u_0(x)$.

Notice  that the function $w$ satisfies the equation
\begin{equation}
  w_t=\epsilon^2 D \Delta w
+ f(v+w, x) - f(v,x) + \epsilon^2 g.
\label {eq3.10}
\end{equation}
One has  $|f(v+w, x) -f(v,x)| < C_2 |w|$ and $g=\Delta
v$. Moreover, $C^2$ regularity of $f$ and the initial data $u_0(x)$ imply
that $|g| < C_1$.

Let us apply the comparison lemma to $w^+_t =  C_2 w^+ + C_1
\epsilon^2$, $w^+(0)=0$, $w^-(t)=-w^+(t)$.
From the relation
$w^+(t)=C_1 {C_2}^{-1}
\epsilon^2(\exp(C_2t)-1)$ it follows that $|w(x,t)| \leq C_3
\epsilon^{2-C_2 a}$ for $t \leq t_\epsilon = -a \log \epsilon$
for positive  $a$ such that $C_2 a < 2$.

Now, using the spectral properties of the matrix $M(x)$ and the
attractive nature of $\phi(x)$, we have
\begin{equation}
  |v(x,t) - \phi(x)| < B \exp(-bt), t> T_0
\label{eq3.11}
  \end{equation}
  that holds uniformly for any $x \in \Omega$.

At $t=t_\epsilon$, one has $|v(x,t_\epsilon)-\phi(x)| \leq \delta
\epsilon^{ba}$, hence $|u_\epsilon(t_\epsilon,x)-\phi(x)| \leq C_4
\epsilon^s$, where $\min(2-C_2a,ba) > s > 0$.

{\bf Part II}. Let us define $\bar w(x,t)$ by
$u_\epsilon(x,t) = \bar w(x,t) + \phi(x) $. Then $\bar w$ satisfies the
equation:
\begin{equation}
\bar w_t = \epsilon^2 D \Delta \bar w + M(x)
\bar w + \epsilon^2 \Delta \phi +
h(\bar w, x,t),
\label{eq3.12}
\end{equation}
where $|h| \le  C_{5} |\bar w|^2$.

From the $C^2$ regularity of $f$ one obtains $|\Delta \phi| < C_6$. Let us
choose $w^+$ to satisfy the equation $$ w_t^+ = -b w^+ + C_5 (w^+)^2
+ C_6 \epsilon^2,
$$
$w^+(t_\epsilon) = |u_\epsilon(x,t_\epsilon) - \phi(x)|$, and
$w^-(t)=-w^+(t)$. Using the strong stability assumption on $M$, we
find that $\max_i \sup_{\xi \in E^i_+(w^+,w^-)} \sum_j M_{ij} \xi_j
\leq -b w^+$. This fact
 ensures that the function $w^+$ satisfies  Lemma 3.1.
The function $w^+$ is less than
 $c{\epsilon^s}$ at $t=t_\epsilon$
and $|w^+|<  c{\epsilon^{s}}$ for all $t \geq
t_\epsilon$.
Thus $|u_\epsilon(x,t) - \phi(x)| < c{\epsilon^{s}}, t
\geq t_\epsilon $. Together with estimate (3.11) this proves the first
inequality (3.7). To prove (3.9), let us observe that
estimate (3.7) can be improved as follows.  Since $w^{+} <
c\epsilon^{s}$, we have
$$
w_t^+ <  -b w^+ + C_7 \epsilon^{s'} w^+ + C_6 \epsilon^2 <
-\frac{b}{2} w^+ + C_6 \epsilon^2.
$$
This implies that
$$
  w_+ <
\bar w(t_{\epsilon}) \exp(- \frac{b}{2})(t -t_{\epsilon}) +
2C_1b^{-1} \epsilon^2 (1 -\exp(- \frac{b}{2})(t -t_{\epsilon})).
$$
This completes the proof.
\vspace{0.2cm}

If $d_1=d_2=... =d_n=d$, this result can be improved. In this case
the strong linear stability condition on $M(x)$ can be weakened. It
is sufficient to suppose that the spectrum of $M(x)$ lies in the
left half-plane for each $x$. More precisely, let us suppose that
there holds the following

{\bf Weak linear stability assumption:}  {\em For each fixed $x$ the
solution of the linear evolution equation
$$
     v_t =M(x) v,
$$
satisfies the estimate
\begin{equation}
     |v(t)| \le  |v(0)| \exp(-\sigma t), \quad \sigma > 0
\label{eq3.13}
\end{equation}
  where $\sigma$ is independent of $x$}.
\vspace{0.2cm}

{\bf Theorem 3.2}. {\em  Suppose that the conditions of Theorem 3.1,
where the strong stability condition is replaced by the weak
stability assumption, hold. Moreover, let the diffusion coefficients
be equal
\begin{equation}
\label{eq3.14}
  d_1= ... =d_n=d \epsilon^2 > 0.
\end{equation}
Then,
 for sufficiently small $\epsilon$, the solution $u^{\epsilon}$
of  problem  (\ref{eq2.1}), (2.2) exists for all $t > 0$, stays close to
a solution of shorted system (2.8) and satisfies the
estimates (3.7), (3.9).}

{\bf Proof}. The first part of the proof repeats,
without any changes, the proof of Theorem 3.1 (in fact, this part
 uses no properties of $M(x)$ excepting for (3.14)).

The second part must be modified and uses now the weak stability
assumption and  condition (3.14).

Let us consider again the equation
\begin{equation} \label{eq3.15}
w_t = \epsilon^2 d \Delta w + M(x) w + \epsilon^2 g +
h(w, x, t),
\end{equation}
where $g=\Delta \phi$, $|h | < c |w|^2$. To estimate the solutions
$w$ of this equation, we introduce the matrix $W(x)$ of size $n
\times n$  defined as follows \cite{DalKr}
$$
   W(x)=\int_0^{\infty} \exp(M(x)^{\dagger}t) \exp(M(x) t) dt.
$$
  This matrix is correctly defined since estimate (3.13) holds. The matrix
$W$ is symmetric and positively defined. Let $(u, v)$ denote the
inner scalar product in ${\bf R}^n$ and $|u|$ is the norm. Then,
$$
   (Wu, u)=\int_0^{\infty} |\exp(M(x) t)u|^2 dt \leq \rho|u|^2,
   \quad \rho = \frac{1}{2\sigma} > 0 $$
and the norms $|u|$ and $(Wu, u)^{1/2}$ are equivalents.

  Moreover, let us notice that  the definition of $W$ entails
\cite{DalKr}
$$
      W M + M^{\dagger} W =-I.
$$
  Let us define now a scalar function $R(x,t)$ by
$R^2 = (W w, w)$ and let us calculate the time derivative
 of this function for solutions $w$ of equation (3.15).
We obtain
$$
  \frac{1}{2} (R^2)_t = d\epsilon^ 2 [(W\Delta w, w) + (W w, \Delta
  w)]
+ (W(h + \epsilon^2 g), w) + (w, W(h + \epsilon^2g)) +
$$
$$
+ (WMw, w)
 + (Ww,  Mw).
$$
  Notice that
$$
  (WMw, w)  +
   (Ww,  Mw)= -|w|^2.
$$
Furthermore,
$$
|(W(h + \epsilon^2 g), w) + (w, W(h + \epsilon^2g)| \le
c_0 R^3 + c_1\epsilon^2 R.
$$
Let us consider the term $Y=(W\Delta w, w) + (W w, \Delta w)$.
This term can be represented as
$$
  Y= \Delta R^2  - 2(W\nabla w, \nabla w) -
2(\nabla W \nabla w, w) - 2(\nabla W w, \nabla w).
$$
Let us notice that
$$
  |(\nabla W \nabla w, w) +
  (\nabla W w, \nabla w)| \le c_{2}(\mu |\nabla w|^2  + \mu^{-1}|w|^2).
$$
Let us choose $\mu$ such that $c_2 \mu < \rho$. Then we find that
$$
   Y \le \Delta R^2 + c_2\mu^{-1} |w|^2 \le \Delta R^2 +
c_3 R^2.
$$
  As a result, one  obtains the differential inequality
$$
  \frac{1}{2} (R^2)_t \le \epsilon^ 2 d \Delta R^2 + c_4\epsilon^2 R
+ c_5 R^3 - R^2 \rho^{-1} + \epsilon^2 c_6 R^2,  \quad t \geq
t_{\epsilon},
$$
  where
$R(t_{\epsilon})= c_7 \epsilon^s,  \ s > 0$. By the standard scalar
comparison principle \cite{Smoller} this differential
inequality implies that
$$
   R < c_8 \epsilon^{s'}, \quad s' > 0.
$$
  The theorem is proved.

To conclude this section let us notice  that it is difficult to
extend Theorem 3.1 to the case, where initial data lie in a
attraction basin of a limit cycle or of a chaotic attractor. For the
limit cycle case, there are formal asymptotic solutions, however,
justification of these solutions is
 a difficult problem. Kuramoto \cite{Kuramoto} has proposed the following
asymptotic approach for the case $f=f(u)$. Supposing that shorted
system (2.8) has a limit cycle solution $\phi( t)$, let us set
\begin{equation}
u^{\epsilon} = \phi( t+ \theta(x,  \tau)) + w,
\end{equation}
where $w$ is a small correction, $\tau=\epsilon^2 t$ is a rescaling
time, $\theta$ is an unknown phase. Then, after some formal
manipulations, one shows that the phase $\theta$ satisfies a
nonlinear diffusion (Burgers) equation which in certain cases can be
resolved analytically.

The proof of Theorem 3.1 shows that, in this case, the solution
$u^{\epsilon}$ attains at a small neighborhood of the limit cycle
within the time $\Ord{-\log \epsilon}$. However, a priori estimates
do not work for all times. Furthermore, phase diffusion phenomena
suggest that, in this case, one can expect a diffusion dependent
patterning. Although the amplitude perturbation $w$ is
small, the phase perturbation $\theta$ is not small and it cannot be found
by the shorted system.

Theorem 3.1 can be extended to the non-autonomous case, where $f$
depends slowly on time but we omit these details.

\section{Applications to genetic circuits}

Diffusion neutral patterning is believed to be the main mechanism in
the morphogenesis of the fruit fly  \cite{Wolpert}. We shall come
back at the end of this section to the correctness of this point of
view but let us first investigate its consequences here. In the
fruit fly embryo, early development just after fecundation consists
in antero-posterior differentiation of body segments. This
patterning is essentially one-dimensional and it results from the
interactions between maternal antero-posterior inducing morphogen
gradients and a set of induced segmentation (gap) genes. The
one-dimensional patterning dynamics has been studied in
\cite{Rein1}, where the gene circuit model is used (see section 2).

Let us consider how our approach works for the genetic circuits
defined by equations (2.1), (2.4).

To begin with, let us notice that in the absence of diffusion the
most general one-dimensional patterning dynamics is defined by a
system of differential equations parametrized by the position $x \in
{\cal K} \subset {\Bbb R}$, where ${\cal K}$ is a compact interval.
More precisely, the patterning dynamics is given by the functions
$q_s(t,x)$ that satisfy the system:

\begin{equation}
 \frac{dq_s}{dt}=Q_s(q,x), \quad q=(q_1,q_2,..., q_p)
\label{eq4.1}
 \end{equation}
 defined on the unit ball $B_p \subset {\bf R}^p$ of dimension
$p$, where the vector field $Q \in C^1(B_p \times {\cal K})$ and at
the boundary this field is directed inward to $B_p$. Then for each
$x$ equations (4.1) define a global semiflow on $B_p$.

We say that a property of  dynamics (4.1) is $C^1$ structurally
stable, if each perturbed system (4.1)
\begin{equation}
 \frac{dq_s}{dt}=Q_s(q,x) +\tilde Q(q,x),
\label{eq4.2}
\end{equation}
such that $|\tilde Q|_{C^1(B_p \times {\cal K})} < \delta$ has the
same property if $\delta$ is sufficiently small.

For  example, the property to have a hyperbolic equilibrium is $C^1$
structurally stable (hyperbolicity means that the matrix of
linearization at the equilibrium has no purely imaginary
eigenvalues). Generally, the property to have an invariant
hyperbolic set with a given topological structure is $C^1$
structurally stable (due to the theorem on the persistence of
hyperbolic sets \cite{Sm,Ru}). The property to have a locally
attracting hyperbolic set is also $C^1$ structurally stable.

Gene circuit models lead to the following shorted
equations:

\begin{equation}
(u_i)_t = \sigma_{\alpha} (\sum_{j=1}^n K_{ij} u_j + \sum_{k=1}^p
J_{ik} m_k(x) - h_i) -\lambda_i u_i.
\label{eq4.3}
\end{equation}

The following theorem expresses the fact that gene circuits models
are sufficiently general as models for diffusion neutral patterning.
\vspace{0.2cm}

{\bf Theorem 4.1}. {\em For each $p$, any system (4.1) and any
$\delta
> 0$ there exists such a choice of parameters $n, K, m, h$ of the
genetic circuit (4.3) such that the dynamics (4.3) has a globally
attracting locally invariant manifold $M_p$ of dimension $p$, for
any $x \in {\cal K}$. This manifold is diffeomorphic to the $p$
-dimensional unit ball $B_p$ and the corresponding reduced dynamics
is defined by the weakly perturbed system (4.2),
 where the
perturbation satisfies $|\tilde Q|_{C^1(B_p \times {\cal K})} <
\delta$.


Therefore, the dynamics of  shorted genetic equations   (4.3)
can have any $C^1$ structurally stable properties
depending on a choice
of
 $n, K, m, h$.


}

{\bf Proof}.  This theorem is a consequence of results in
\cite{Vak5, Vak6}.

{\bf Remark}. A too strong diffusion or a diffusion dependent patterning
would impose restrictions on the structurally stable properties that can
be  reproduced by systems (2.1) with nonlinearities (2.5). We have already
mentioned the phase diffusion phenomena for limit cycles. Also, for
large diffusion coefficients the derivatives of the point attractors
$u^{\epsilon}_\infty(x) = \lim_{t \to \infty} u^{\epsilon}(x,t) $
of system (\ref{eq2.1})
 with respect to $x$ cannot take large values.
Finally, diffusion may destabilize point attractors via the Turing
mechanism. Turing instability is diffusion dependent patterning and
is excluded by conditions (\ref{eq3.5}) or (\ref{eq3.14}).



Providing that all the hypothesis of Theorem 3.1 or 3.2  are fulfilled we
can say that the introduction of diffusion has a small effect on
patterning; patterning is diffusion neutral.

The main hypothesis of Theorems 3.1 and 3.2 refers to the
structure of the attractors of the shorted equation that we shall
analyze below.

The steady states of the shorted equation are solutions of

\begin{equation}
u_i = \lambda_i^{-1} \sigma_\alpha (\sum_{j=1}^n K_{ij} u_j +
\sum_{k=1}^p  J_{ik} m_k(x) - h_i) = G_i(u)
\end{equation}

For large $\lambda_i$, the application $u \to G(u)$ is a contraction
map and thus an unique stable rest point $u_0$ exists. In this case
patterning is diffusion neutral. For smaller $\lambda_i$, it is
possible to have coexistence of several point attractors. A
necessary criterium for multi-stationarity is the existence of a
positive loop in the interaction graph defined by the matrix $K$
\cite{soule}. For competitive gene circuits ($K_{ij} \leq 0$ for all
$i \ne j$) such as the gap gene circuit of the fruit fly a positive
loop means a loop made of an even number of interactions. As pointed
out by Smale \cite{smale-mathbio} competitive systems in dimension
$n$ can have any dynamics that is possible in $n-1$ dimensions.
Chaotic attractors could be expected for $n \geq 4$ and limit cycles
for $n \geq 3$. Competitive gene circuits for  $n=2$ have only point
attractors  and for $n=3$ all attractors are either limit circles or
points \cite{hirsch83,hirsch90}. Cooperative gene circuits ($K_{ij}
\geq 0$ for all $i \ne j$) are monotone and monotone systems have
particularly simple dynamics: almost all trajectories converge to
point equilibria \cite{ST}. Some gene circuits can be made
cooperative just by a change of coordinates. It is the case of gene
circuits such that any loop in the interaction graph has an even
number of negative interactions.

 Although there are
no general methods for finding attractors, in some special cases
there are algorithms allowing the exhaustive determination of
point attractors. In the following we discuss such a special
situation.

Let us consider the case $\alpha >> 1$ and let us restrict for
simplicity to the case of a single morphogen. Then the steady state
solutions of the shorted equations have the following form as $\alpha \to
\infty$:

\begin{equation}
  u_i^0 =\lambda_i^{-1} s_i(x) + O(\alpha^{-1}), \quad s_i(x) \in \{0, 1\}.
  \label{4.4}
\end{equation}
Locally in $x$ the steady state of the shorted equation are discrete
and are indexed by the set $\{0, 1 \}^n$. They are the solutions of
the following binary programming problem:
\begin{equation}
\sum_{j=1}^n \tilde{K}_{ij} s_j(x) + m(x) > \tilde{h}_i , \quad if \
s_i(x)=1,
\end{equation}
\begin{equation}
\sum_{j=1}^n \tilde{K}_{ij} s_j(x) + m(x) < \tilde{h}_i , \quad if \
s_i(x)=0,
\label{4.5}
\end{equation}
where $\tilde{K}_{ij} = \frac{K_{ij}}{J_{i1}\lambda_j}$,
$\tilde{h}_i = \frac{h_i}{J_{i1}}$, $m(x)=m_1(x)$.

In the fruit fly embryo $m(x)$ is a monotonous function of the
anteroposterior position $x$ ($q=1, x \in {\Bbb R }$). The solution
of the programming problem can be given by specifying for each
possible steady state ${\cal
S}^{(k)}=(s_1^{(k)},s_2^{(k)},\ldots,s_n^{(k)}) \in \{0, 1 \}^n$ the
domain ${\cal I}^{(k)}$ in $x$, where this state exists. If $m(x)$
is monotonous, then all ${\cal I}^{(k)}$ are intervals:
$${\cal
I}^{(k)} =\{ x |  \max_{s_i^{(k)}=1} (\tilde h_i - \sum_j \tilde
K_{ij} s_j^{ (k)} ) < m(x) <   \min_{s_i^{(k)}=0} (\tilde h_i - \sum_j
\tilde K_{ij} s_j^{(k)}) \}.
$$
The pattern consists of segments corresponding to different states
$ {\cal S}^{(k)}  $. If the intervals ${\cal I}^{(k)}$ have
non-overlapping interiors, the interfaces separating segments are
transition layers. Overlapping interval interiors are compatible
with the presence of diffusion layers. As illustrated in Figure 4,
the existence of diffusion layers may ask for extra conditions on
the initial data or may be forced for all initial data.

In general,  the binary programming problem is a very difficult problem
for large numbers of genes.
Nevertheless, early stages of
morphogenesis correspond to a small number of interacting genes and
the complete analysis of the problem is possible. Let us consider
two examples of small complexity:

{\bf Example 1 : $n=1$, one component.}

In the gene circuit model, the one-component case describes the
situation of a zygotic gene $u_1$ which is not regulated by other
zygotic genes $u_j, j\neq 1$, i.e. $K_{1j}=0$ for any $j \neq 1$.

In this case there are two possible steady states, ${\cal
S}^{(1)}=(1)$,${\cal S}^{(2)}=(0)$, existing in the intervals
$I^{(1)} = \{x | m(x) >  \tilde{h}_1 - \tilde{K}_{11} \}$ and
$I^{(2)} = \{x | m(x) <  \tilde{h}_1 \}$, respectively. If $K_{11}
= 0$ (gene $1$ has no regulation effect on itself) the two
interval interiors do not overlap. The patterning is diffusion
neutral. The interface between the two states is a transition
layer and for $\epsilon << \alpha^{-1}$ its width is of order
$\Ord{\alpha^{-1}}$ (see Fig.\ref{fig1}~a). This one-component case is
the one usually used by biologists to illustrate Wolpert's
positional information mechanisms \cite{Wolpert} and the diffusion
neutrality hypothesis is entirely justified here.

The case $K_{11} < 0$ (gene 1 inhibits itself) needs special
treatment. Indeed, in this case there is a gap between the
intervals $I^{(1)}$ and $I^{(2)}$ on which the boolean programming
problem has no solution at all. This phenomenon is due to the
singular character of the limit $\alpha \to \infty$ and disappears
for finite $\alpha$. Let us consider the function $f_\alpha (u, m
) = \sigma_\alpha(K_{11} u + J_{11}m - h_1) -\lambda_1 u$. The function
$f$ is monotonic in $m$. The steady state of the shorted equation
satisfies $f_\alpha (u, m )=0$. If $m \in (\tilde{h}_1,\tilde{h}_1
- \tilde{K}_{11})$, then $f_\alpha (0, m )
>0$, $f_\alpha (1/\lambda_1, m ) < 0$ and, therefore, there
is an unique $u_0(m)$, $0 < u_0(m) < 1/\lambda_1$ such that
$f_\alpha (u_0(m), m )=0$. The function $\phi(x)=u_0(m(x))$
describes a smooth transition layer connecting the states
$u_1\Ord{\alpha^{-1}}$ and $u_2 = 1/\lambda_1 + \Ord{\alpha^{-1}} $.
Patterning is diffusion neutral because the branch of attractors
$\phi(x)$ is unique. The pattern contains a  transition layer whose
width does not reduce to zero as $\alpha \to \infty$ (see
Fig.\ref{fig1}~b).

If $K_{11} > 0$ (gene $1$ activates itself) there is an overlap of
the intervals $I^{(1)}$,$I^{(2)}$ and it is possible to have type 2
interfaces (diffusion layers) and diffusion dependent patterning. In
the last section we shall show  that in this case the width of the
interface is controlled by the diffusion coefficient $D$ and that
this width vanishes in the limit $\epsilon \to 0$
Fig.\ref{fig1}~c,d. Fig.\ref{fig1}~c illustrates a situation when
the interval $I^{(2)}$ extends over the entire patterning domain.
Depending on initial data patterning can be diffusion neutral or
dependent. Initial data presented in Fig.\ref{fig1}~c lead to a
diffusion layer. Another choice of initial data $u_0(x)=1$ (not
shown in figure) would lead to a diffusion neutral steady state that
has no interface. The case Fig.\ref{fig1}~d does not leave this
choice: irrespectively of data the pattern is forced to contain a
diffusion layer and is diffusion dependent.

{\bf Example 2: $n=2$, two components.}

Let us suppose that $\tilde{K}_{ii}=0, i=1,2, \tilde{K}_{12}\leq
0, \tilde{K}_{21}\leq 0$. In this case there are four possible
steady states, ${\cal S}^{(1)}=(1,1)$,${\cal
S}^{(2)}=(0,0)$,${\cal S}^{(3)}=(1,0)$,${\cal S}^{(4)}=(0,1)$. The
corresponding existence intervals are $I^{(1)} = \{x | m(x) >
\max( \tilde{h}_1  - \tilde{K}_{12} , \tilde{h}_2  -
\tilde{K}_{21} )  \}$, $I^{(2)} = \{x | m(x) <  \min( \tilde{h}_1
, \tilde{h}_2   )  \}$, $I^{(3)} = \{x | \tilde{h}_1 < m(x) <
\tilde{h}_2  - \tilde{K}_{21}  \}$, $I^{(4)} = \{x | \tilde{h}_2 <
m(x) < \tilde{h}_1  - \tilde{K}_{12} \}$. Notice that $I^{(3)}$
and $I^{(4)}$ can overlap if the following condition is satisfied:

\begin{equation}
\tilde{h}_1 \leq \tilde{h}_2  \leq \tilde{h}_1  - \tilde{K}_{12},
\quad \text{or} \quad \tilde{h}_2 \leq \tilde{h}_1  \leq \tilde{h}_2
- \tilde{K}_{21}
\label{4.8}
\end{equation}

In general, existence of a diffusion layer connecting ${\cal
S}^{(3)}$ and ${\cal S}^{(4)}$ depends on the initial data and on
 condition (4.8).

A diffusion layer  always exists if an
extremity of $I^{(3)}$ belongs to the interior of $I^{(4)}$, or
reciprocally if an extremity of $I^{(4)}$ belongs to the interior of
$I^{(3)}$. This condition corresponds to strict inequalities in
(4.8).

The interiors of $I^{(1)}$, and of $I^{(2)}$ do not overlap on the
interiors of other intervals, therefore the interfaces separating
the states ${\cal S}^{(1)}$ or ${\cal S}^{(2)}$ from any other
states are transition layers. Contrary to the preceding example,
there are no longer gaps between intervals meaning that for small
diffusion coefficients the widths of all the transition layers are
of the order $\Ord{\alpha^{-1}}$.

 To conclude, in this case diffusion neutrality is
justified for interfaces involving ${\cal S}^{(1)}$ or ${\cal
S}^{(2)}$ but it is not always justified for interfaces separating
${\cal S}^{(3)}$ from  ${\cal S}^{(4)}$.

\section{Diffusion dependent patterning}

If condition (3.3) of Theorem 3.1 is invalid, i.e., if initial data
$u_0(x)$ lie in  attraction basins of different attractors of the
shorted system
 for different $x$, then the estimates of the previous section hold only on
 a time interval of order $O(-\log \epsilon)$.
The main reason for that is the following. In this case the
derivatives of the solution $v (x, t)$ of the shorted equations with
respect to $x$ increase unboundedly at some fixed positions $x =
q_0$ as $t \to \infty$. In the presence of diffusion, one can expect
that there exist diffusion layers connecting different attractors of
the shorted system. These layers are mobile. They eventually reach
equilibrium positions which are generally different from the initial
positions $q_0$.

To understand this situation, let us remind some basic definitions
from the theory of finite dimensional dynamical systems \cite{Ru}.
Let us consider the time autonomous shorted system (2.8) ($f=f(v,
x)$) for a fixed $x\in \Omega$ and let us suppose that the uniform
dissipativity condition holds in a ball $B(R_0)$ in ${\bf R}^n$. Let
us recall that $u_0(x) \in B(R_0)$ is wandering if there exists a
neighborhood $U(u_0)$ of $u_0$ and a time $T_0 >0$ such that for all
$t > T_0$ the trajectory $v(t, x, u_0(x))$ starting from $u_0(x)$
 does
not intersect $U(u_0)$: $v(t, x, u_0(x)) \notin U(u_0), \ t > T_0$.
In the theory of finite dimensional dynamical systems, the set of
non wandering points play a key role
 \cite{Ru}. It is a closed invariant set
which contains the $\omega$-limit  sets of all trajectories ($w \in
{\bf R}^n$ is the $\omega$- limit set of the trajectory  $v(t, x,
u_0(x))$ if and only if there is a sequence $t_k \to \infty $  such
that $v(t_k, x, u_0(x)) \to w $  as  $k \to \infty$). In our case
these sets depend on $x$, since $x$ plays the role of a parameter.
Typically, the non wandering set $\Theta_x$ consists of some
connected components ${A^i}_x$. Some components are local attractors
(i.e., $A^i_x$ attracts all trajectories starting in an open
neighborhood of $A^i_x$), others are saddle sets and repellers.
Intuitively, initial interface position   can be associated with $x$
such that the corresponding trajectory $v(x, t, u_0(x))$   goes to a
 saddle point or hangs in a fixed repeller as $t \to \infty$ \cite{MOLV}.


However, a rigorous mathematical analysis is nontrivial even in the
simplest cases. To illustrate, let us consider two simple examples.
In these examples we set periodic boundary conditions 
in $x \in \Omega$, supposing that $\Omega$ is a box.

{\bf Example 1}. Let us consider a system of Allen- Cahn's type:
\begin{equation}
  u_t=\epsilon ^2 \Delta u     + a(x) (u -u^3), \quad u \in \R,
\label{AC47}
\end{equation}
where $a(x) > 0, x\in \Omega \subset \R^q$. The nonwandering set of
the shorted system is $\{ -1, 0, 1 \}$,  where  $1, -1$ are local
attractors and $0$ is a saddle point. If $u_0(x) > 0$ for all $x \in
\Omega$ or if $u_0(x) < 0$ for all $x \in \Omega$, one can apply
Theorem 3.1 or 3.2 and $u(t, x) \to \pm 1 + \Ord{\epsilon^s}$ as $t
\to \infty$, respectively.

If there are points, where $u_0(x)=0$, we  can expect a more
complicated behavior. The analysis \cite{MOLV} shows that the
gradient of $u(x, t)$ increases to $\infty$ at the set $S=\{x:  \
u_0(x)=0 \}$. Fife \cite{FifeHsi} described in detail the
development of the inner layer in the one-dimensional case ($q=1$)
and when the set $S$ contains a single point.

For arbitrary $u_0(x)$, the set $S$ can be a rather complicated,
fractal set.
 If  $u_0 \in C^{\infty}$ and if the rank of the
 derivative $\text{d} u_0(x)$ is $r, 0 \leq r \leq q$
  then  by the implicit function theorem
$S$ is a smooth sub-manifold of $\Omega$ of co-dimension $r$. Like
in \cite{MOLV}, we use the concept of "generic situation", which is
standard in differential topology \cite{HirschI}. If the initial
data $u_0$ are smooth and generic, $\text{d} u_0(x)$ has full rank
$r=1$ (which is the dimension of the u-phase space), hence $codim
S=1$. This means that, within times $t$ of order $|\log \epsilon|$,
one obtains an interface appearance at some hypersurfaces in
$\Omega$. These hypersurfaces are initial positions for interface
evolution for $t >> O(|\log \epsilon|)$. \vspace{0.2cm}



{\bf Example 2}. Let us consider a system of Ginzburg -Landau's type:
$$
  u_t=\epsilon ^2 \Delta u     + a(x) (u -|u|^2 u),
$$
where $u$ is an unknown complex valued function,
 $a(x) > 0$ in $\Omega$.

The non wandering set of the shorted system is the union of the
attracting limit cycle  $|u|=1$ and the repeller $\{0 \}$. The set
$S$ is now, for generic smooth $u_0(x)$, a submanifold of
codimension $2$. This means that if $dim \ \Omega=1$, generically,
we have no singularity growth, if $dim \ \Omega=2$ one obtains
vortices localized at some points and if $dim \ \Omega=3$ we have
vortex lines \cite{MOLV}.




Let us now formulate
 a result identifying the diffusion
dependent patterning situation considered in the above examples.

Let us denote by $v(t, x, u_0(x)) $
the solution of  the time autonomous shorted system (2.8) with the
initial data $u_0(x)$.

Let us formulate some conditions on  dynamics of shorted system
(2.8).  Let us notice that among the components of $\Theta_x$ there
exists a component $A_x \subset \Theta_x$  attracting a set $B_x$ of
points $v$ with a non-empty interior. We denote by $A^{\prime}_x$
the complement of $A_x$ in $\Theta_x$.  In the general case, the
dependence of $\Theta_x$ and $A_x$ on the parameter $x$ may be very
nontrivial. To simplify, we impose the following
 conditions: \vspace{0.2cm}

{\bf C1}  {\em For each $x$ the nonwandering set $\Theta_x$ of
(2.8) contains a set $A_x$ attracting for an open neighborhood $V_x$,
$A_x \subset V_x$.
This means  that all the trajectories
$v(t, x, v_0)$, starting at the points $v_0 \in V_x$
 satisfy the condition:
$$
   dist\{ v(t, x, v_0), A_x \}  \to 0
$$
as $t \to \infty$.
Assume that
$V_x$ continuously depends on $x$, i.e.,
$$
     dist \{ V_x, V_y \} \to 0 \quad
$$
  as $x \to y$.}
\vspace{0.2cm}

{\bf C2} {\em The sets
$A_x$ do not intersect the other components of the nonwandering sets
$\Theta_y$, i.e.,
$$
    A_x \cap A^{\prime}_y = \emptyset
$$
for any $x, y$.
}
\vspace{0.2cm}

The conditions on the attractors are fulfilled in the examples 1, 2
or more generally when $f(v, x)$ is scalar and has some continuous
non-intersecting branches of roots $v_k(x)$, $k=1,2,...,M$, where
$f_v(v_k(x), x) < 0$, serving as point attractors. It is the case of
the Allen-Cahn model. \vspace{0.2cm}

{\bf Theorem 5.1}. {\em   Consider the time autonomous case
$f=f(u,x), \ f \in C^2$, where
$\Omega$ is a box and we set
the periodic boundary conditions.
Assume that shorted system
(2.8) satisfies the uniform dissipativity condition (\ref{eq3.2}),
and let the initial data satisfy
$$
  |u_0(x) | < R,
$$
  where $R$ is the constant from the uniform dissipativity condition.

Suppose  the shorted system satisfies  conditions {\bf C1,
C2}.
Suppose, moreover, that there exist two distinct points $x_0, x_1 \in
\Omega$ such that the  solutions of the shorted system
satisfy
$$
  v(t, x_0, u_0(x_0)) \to  {A}_{x_0}, \quad
$$
and the $\omega$ -limit set of the trajectory $  v(t, x_1,
u_0(x_1))$ lies in the complement $A^{\prime}_{x_1}$ of $A_{x_1}$.

Then patterning with the initial data $u_0(x)$ is diffusion dependent}.


{\bf Proof}.

Let us suppose that patterning is diffusion neutral.
Let us consider the set $\Omega_V$ consisting of $x \in \Omega$
such that the trajectory $v(t, x, u_0(x))$ enters $V_x$ for some $t=t_0(x)$:
$v(t_0, x, u_0(x))  \in V_x$.
It is clear
$\Omega_V$ is an open set since
$V_x$
 depends continuously on $x$ and trajectories
$v(x, t, u_0(x))$ also depend continuously on $x$
within time bounded intervals.  If $x \in \Omega_V$, the corresponding
trajectory $v(t, x, u_0(x))$ of shorted system (2.8) converges to
$A_x$. Indeed, this trajectory enters $V_x$, and since $V_x$ is contained in
the basin attraction of $A_x$, this trajectory tends to $A_x$ as
$t \to \infty$.

Let
$\overline{\Omega}_V$ be the closure of  $\Omega_V$  in $\Omega$. As
 $\Omega_V$ is open in the connected set $\Omega$, there
exists a point $x_*$ such that $x_* \in \overline{\Omega}_V$ and
$x_* \notin \Omega_V$. Thus, for any $\delta>0$ there exists
$y_{\delta} \in \Omega_V$ such that $dist ( y_{\delta}, x_* ) <
\delta$. Consider now the trajectories
$v(t, y_{\delta},
u_0(y_{\delta}))$
and
the trajectory
$v(t, x_*, u_0(x_*))$.
The first ones converge
to the corresponding sets $A_{y_{\delta}} \subset V_{y_{\delta}}$.
The trajectory
$v(t, x_*, u_0(x_*))$ approaches to a $\omega$ -limit set $C_*$
as $t \to \infty$
which does not intersect $A_{x_*}$. Thus this
$\omega$ -limit set $C_*$ lies in the complement $A^{\prime}_{x_*}$.
 Therefore, this set $C_*$ does not intersect any $A_y$ for
any $y$ (due to assumption {\bf C2}).
Thanks to these limit properties of
the trajectories $v(t, x_*, u_0(x_*))$  and
$v(t, y_{\delta},
u_0(y_{\delta}))$
one concludes that, for sufficiently small $\kappa > 0,
\delta > 0$ there is a time moment $T(\kappa, \delta)$ such that
$$
  |v(t, y_{\delta}, u_0(y_{\delta})) -  v(t, x_*, u_0(x_*))| > \kappa,
\quad t=T(\kappa, \delta).
$$
   Let us choose a small $\epsilon$ such that $r_{\epsilon} < \kappa/2$.
Let us fix such $\kappa, \epsilon$. Then the previous estimate and
the definition of diffusion neutral patterning yield
$$
  |u(t, y_{\delta}) -  u(t, x_*)| > \frac{\kappa}{2},
\quad t=T(\kappa, \delta),
$$
where $dist ( y_{\delta}, x_*) < \delta$.
     This  holds for any $\delta$. Letting $\delta \to 0$
one notices that
$$
        |\nabla u(x_*, T(\kappa, \delta))|  \to \infty  \quad (\delta \to 0).
$$
However,  solutions of our problem (2.1), (2.2) are  a priori
bounded  due to the uniform dissipativity. Therefore, the Schauder a
priori estimate is fulfilled: $|u_x| < c \epsilon^{-1}$
\cite{Kruzkov}. This fact gives us a contradiction with the last
limit relation. The theorem is proved.

Now, let us describe some properties of diffusion layers.
Unfortunately, there is not a general theory on such layers for $n
>1$. The diffusion layers can be described in the one-component case
$n=1$, and for $n>1$ only in the case of gradient systems, or in the
case of monotone systems   \cite{VVA1,VVA2,VVA3}. Let us remind that
gene circuits generate a monotone dynamics if the interaction matrix
$K_{ij}$ satisfies the cooperativity condition $K_{ij} \geq 0, \ i
\ne j$.





It is convenient to analyze the motion of internal layers in two
steps. We shall only deal with the one-component case $n=1$.

{\em I. Internal (diffusion) layer problem in  infinite
homogeneous media}

Let us ignore  temporarily the dependence of $f$ on the space
variable $x$ and on the slow time variable $\tau=\epsilon t$ and
let us suppose that $f$ has the shape in Fig.~\ref{fig1}~c).
This means that $f(u,x,\tau)$ satisfies the following assumptions
(bistability): for each $x,\tau$, the function $f(u,x, \tau)$ has
only three roots $w_0(x,\tau), w_1(x, \tau)$ and $w_{2}(x, \tau))$
such that
\begin{equation}
  f_u(w_0,x, \tau) <0, \quad f_u(w_1, x, \tau) < 0, \quad
  f_u(w_{2}, x, \tau) > 0, \quad w_0 < w_{2} < w_1,
\end{equation}
for any $x, \tau$.

Notice that this assumption is  satisfied for the Allen-Cahn
model. It is also satisfied for gene circuits under some
conditions such as in the situation illustrated in
Fig.~\ref{fig1}~c).

 Furthermore, let us consider that the pattern is infinite, i.e.,
 $x \in (-\infty, +\infty)$.
This leads to the following problem involving the parameters
$q,\tau$:
\begin{equation}
{u}_{t} =    u_{xx} + f(u, q, \tau),
\label{5.1}
\end{equation}
\begin{equation}
 \lim_{x \to -\infty} u(x)=w_0(q, \tau), \quad \lim_{x \to +\infty}
u(x)=w_1(q, \tau).
\label{5.2}
\end{equation}

Denote by $\Phi(u,x,\tau)$ the
primitive $\Phi(u,x, \tau) = \int_{w_0}^u f(s,x, \tau) ds$.

The well known result for this problem can be formulated as follows.
\vspace{0.2cm}

{\bf Proposition 5.2} {\em The internal (diffusion) layer problem
(\ref{5.1}), (\ref{5.2}) has traveling wave solutions $u=\psi(x-
V(q,\tau)t, q, \tau)$, where the velocity $V$ is a functional of $f$
(and thus depends on the parameters $q,\tau$).
 The velocity $V(q,\tau)=0$
if $\Phi(w_0,q, \tau)=\Phi(w_1, q, \tau)$. The function
$\psi$  is increasing in $x$
for any fixed $q, \tau$ and satisfies the following exponential
estimates
\begin{equation}
 |\psi(z) - w_0| < C \exp(b_0 z),  \quad z < 0,
\end{equation}
\begin{equation}
 |\psi(z) - w_1| < C \exp(-b_1 z),  \quad z > 0,
\end{equation}
and
\begin{equation}
  |\psi_z| < C \exp(-b|z|), \quad b=\min\{b_0, b_1\},
\end{equation}
where $C, b_i, b$ are positive constants depending on $q, \tau$.
}


{\bf Proof. } Below we sometimes omit dependence on $q, \tau$ in
notation. The function $\psi (\xi,q,\tau)$  satisfies the equation
\begin{equation}
\psi_{\xi \xi} + V {\psi}_{\xi} + f(\psi,q, \tau) =0. \label{twave}
\end{equation}

The existence of a solution of equation (\ref{twave})
 satisfying exponential estimates
 is well known, for example, \cite{FifeMac, VVA2}.
Multiplying equation (\ref{twave}) by ${\psi}_{\xi}$, replacing
$f={\Phi}_{\psi}$ and integrating one obtains :
\begin{equation}
{\Phi(w_0)- \Phi(w_1)}=\frac{V}{2} {\int_{-\infty}^{\infty}
\left({\psi}_{\xi}\right)^2 d \xi}.
 \label{eqc}
\end{equation}
This equation  shows that the velocity $V=0$ and thus the traveling
wave is a stationary interface if $\Phi(w_0)=\Phi(w_1)$.




{\bf Remark 1}. If $V=0$, equation (5.8) has a first integral $
\frac{1}{2}{\psi}_{\xi}^2 + \Phi(\psi) = \Phi(w_0)$. This fact
allows us to calculate the integral
$$
{\int_{-\infty}^{\infty} \left({\psi}_{\xi}\right)^2 d \xi} =
\sqrt{2} \int_{w_0}^{w_1} \sqrt{\Phi(w_0)-\Phi(\psi)} d\psi.
$$

Thus, one obtains an approximate formula valid
 for small interface velocities:
\begin{equation} V \approx \frac{\sqrt{2} [
\Phi(w_0) - \Phi(w_1)]}{\int_{w_0}^{w_1} \sqrt{\Phi(w_0)-\Phi(\psi)}
d\psi } \label{velocity}
\end{equation}

{\bf Remark 2}. Constants $b_i$ can be computed by a linearization
of equation (5.3) at $w_0$ and $w_1$ respectively. We obtain
\begin{equation}
-V(q,\tau) b_i = b_i^2 + f_u(w_i, q,\tau)
\label{5.3}
\end{equation}

{\bf Remark 3}. This result can be generalized  for monotone
systems, see \cite{VVA1,VVA2,VVA3}. However, even for weakly
non-monotone systems this result is not correct: there are waves
with a very complex structure, whose wave front profiles change with
position and time \cite{VakVol2}. For these examples the wave front
cannot be presented as a function of $x-Vt$ and Proposition 5.2 does
not hold; we shall not consider these questions here.

{\bf Example 1}.

Let us consider the interface properties for a genetic circuit with
a single zygotic gene and a single morphogen ($n=1, K_{11}=1,
m(q)=J_{11} m_1(q)
> 0,\, \lambda=\lambda_1, \, h= h_1$). Supposing that $\alpha$
(sharpness) is a large parameter, then for positions such that $h -
1/\lambda  + \Ord{\alpha^{-1} }  < m(q) < h + \Ord{\alpha^{-1} } $
we obtain two stable stationary solutions ($ u_- = \Ord{\alpha^{-1}
}, u_+=\lambda^{-1}+ \Ord{\alpha^{-1} }$ and one unstable $u_s$
defined by the relations $u_s=h - m + \alpha^{-1} v$,
$\sigma(v)=\lambda (h - m + \alpha^{-1} v), \ |v| < C$.


In this case one has $\Phi(u) = (u+m-h) \text{H}(u+m-h) -\lambda u^2/2 +
\Ord{\alpha^{-1} }$, where $\text{H}$ is the Heaviside step
function. By this relation and  (\ref{eqc}) one finds
\begin{equation}
    V(q) =  \rho(q)[\frac{1}{2\lambda} -h+m(q)] +\Ord{\alpha^{-1}},
\label{5.4}
\end{equation}
where $\rho >0$ for all $q$. Using (\ref{velocity}) one obtains the
approximation $\rho(q) \approx 2 \lambda \sqrt{2 \lambda} $ valid
for small velocities. We have a stationary interface if $m(q) = h -
\frac{1}{2\lambda} + \Ord{\alpha^{-1}}$.

{\bf Example 2}.

The shorted equation of the Allen-Cahn model  has two attractor
nodes $u_0,u_1$ and a saddle $u_2$. The values $u_0,u_1$ are
 the two maxima and $u_2$
is the minimum of the fourth order polynomial potential $\Phi(u)$.
In this case the traveling wave velocity is
\begin{equation}
V(q,\tau) =  \sqrt{2 A(q,\tau)} [ u_0(q,\tau) + u_1(q,\tau) - 2
u_2(q,\tau) ].
\end{equation}


There exist thus stationary interfaces, where the saddle is precisely at half
distance between the two nodes:
\begin{equation}
u_2(q,\tau) = [u_0(q,\tau) + u_1(q,\tau) ]/ 2.
\end{equation}
\vspace{0.2cm}

{\em II. Internal (diffusion) layer problem in heterogeneous
media}

After establishing Proposition 5.2, we can get a formal asymptotic
solution of the interface motion problem. Let us set
\begin{equation}
  u_{as}=\psi(\theta(x,t, \epsilon), q, \tau) + O(\epsilon),
\end{equation}
  where
\begin{equation}
  \theta=\epsilon^{-1}(x-q(t))
\end{equation}
and where a unknown function $q(t)$  determines the localization
of the narrow interface. By substituting the expression for $u_{as}$ in
equation (\ref{eq2.1}) one observes that the terms of the principal
order $O(1)$ vanish under the condition
\begin{equation}
  \frac{dq}{dt}=\epsilon V(q, \epsilon t).
\label{asv}
\end{equation}
One can therefore expect that this equation describes the
interface (diffusion layer) propagation. If $f$ depends only on
$x$, there exist stationary solutions of equation (\ref{asv}).
   Indeed, if $q_1, q_2,...,q_l$ are roots of the function $V(q)$ then
the diffusion layer is immobile in any one of the  positions
$q_j$. These equilibrium positions can be stable or unstable. It
is easy to see that if the derivative $V^{_j}_q > 0$, then the
corresponding position is unstable, and if $V^{_j}_q < 0$, the
position is stable. In this situation there are stable stationary
solutions with diffusion layers.

After this heuristic reasoning let us formulate  the main result of this
section.



Let us assume  that the function $f$ satisfies {\it uniform dissipativity}
condition, namely, there exists positive constants
$M_1, M_2$ uniform in $x, \tau$ such that
\begin{equation}
  f(u, x, \tau) < 0, \quad u > M_1,
\end{equation}
\begin{equation}
  f(u, x, \tau) > 0, \quad u <  -M_2.
\end{equation}

This assumption ensures the existence of an unique smooth solution
of (\ref{eq2.1}) (see \cite{Henry}). Notice that the uniform
dissipativity  condition always holds for one-component genetic
circuits. It also holds for the Allen-Cahn model if the functions
$u_0(q,\tau),u_1(q,\tau)$ are uniformly bounded. \vspace{0.2cm}


{\bf  Theorem 5.3 (on the interface motion). } {\em Let us consider
the reaction-diffusion equation
\begin{equation}
\label{eqRD}
u_t=\epsilon^2  u_{xx} + f(u,x, \epsilon t), \quad
u(x,0)=u_0(x),
\end{equation}
where $x \in [0,1]$,
under the  zero Neumann  boundary conditions
$$
  u_x(0, t)=u_x(1, t)=0.
$$
Assume  $f \in C^2$. Let us suppose that the
assumptions of Proposition 5.2 hold and
$$
  w_1(x,\tau) -   w_0(x,\tau)  > \delta_0 > 0,
$$
\begin{equation}
  \sup_{x,\tau} f_u(w_i, x, \tau) = -\mu_i < 0, \quad i=0,1,
\label{df}
\end{equation}
where $x \in [0,1]$, $\tau > 0$. Moreover, assume that
\begin{equation}
  ({w_0)}_x(0, \tau) < 0, \quad  {(w_1)}_x(1, \tau) > 0.
\label{condw}
\end{equation}

{\bf I}. Let $\psi(\xi,q,\tau)$ be an interface solution of eq. (5.8)
with asymptotic boundary conditions $\psi(-\infty,q,\tau)=w_0(q,\tau)$,
$\psi(\infty,q,\tau)=w_1(q,\tau)$ (see Prop.5.2).
Suppose that the initial data are
sufficiently close to this interface. More precisely,
\begin{equation}
 |u_0(x) - \psi(\epsilon^{-1} (x-q_0), q_0, 0)| < \bar c_0\epsilon^{s},
\quad s \in (0,1),
\label{initcond}
\end{equation}
Then, if $\epsilon$ is small enough,  the solution of the problem
has the interface form
\begin{equation}
u=\psi(\epsilon^{-1} (x-q(t)), q(t), \epsilon t) + v,
\label{travelw}
\end{equation}
where the correction  $v$ satisfies the estimate
\begin{equation}
  |v| <  C\epsilon^{s},    \quad C > \bar c_0 > 0
\end{equation}
and the time evolution
of the interface position $q$ is defined by the differential equation
\begin{equation}
  \frac{dq}{dt}= \epsilon (V(q,\epsilon t) + R(q, \epsilon, t)), \quad q(0)  =  q_0,
\label{evinf}
\end{equation}
where
$$
   |R| < c'\epsilon^{s_1}, \quad  s_1 > 0.
$$
Equation (\ref{evinf}) holds while
$q > c\epsilon^{s_2}$ and
$q < 1 - c\epsilon^{s_2}$, where $s_2 \in (0,1)$.

{\bf II}. Consider the time-independent case: $f=f(u,x)$. If there
is a point $q_*$ such that
\begin{equation}
  V(q_*)=0,  \quad V^{\prime} (q_*) < 0
\end{equation}
  then there is a stationary, well localized at $x=q_*$
interface solution  tending to a step-like function as $\epsilon \to 0$.}



{\bf Remark 1}. The assertion {\bf II} is a consequence of results
obtained by Fife \cite {Fife}, who also considered the time
evolution of interfaces in the case $f=f(u,x)$ \cite{FifeHsi}.

{\bf Remark 2}.  Condition (\ref{condw}) is technical and simplify
estimates.

{\bf Remark 3}.  Condition (\ref{df}) yields that the functions
$b_i, i=0,1$ defined by Eqs. 5.5, 5.6 satisfy;
$$
  b_i(x, \tau) > \tilde b_i > 0, \quad i=0,1,
$$
 for all $x \in [0,1], \ \tau > 0$.

{\bf Proof of part I}.
 Below some key estimates can be
simplified if $w_0, w_1$ are independent of $x, \tau$. We reduce the
general situation to this case by introducing a new
variable $\tilde u$:
\begin{equation}
  u =\beta(x, \tau) \tilde u + w_0(x,\tau), \quad
\beta =w_1(x, \tau) -w_0(x, \tau),
\end{equation}
where $\tau=\epsilon t$.

 For the new unknown function $\tilde u$ the corresponding values
$\tilde w_i$ are $0,1$, respectively.  Moreover, the boundary
conditions take the following form
\begin{equation}
\beta \tilde u_x(0,t)=- {w_0}_x(0, \tau) -\beta_x(0, \tau) \tilde u(0,t),
\end{equation}
\begin{equation}
\beta \tilde u_x(1,t)=- {w_0}_x(1, \tau) -\beta_x(1, \tau) \tilde u(1,t).
\end{equation}
Making  the change $u \to \tilde u$
one obtains the equation
\begin{equation}
\tilde u_t =\epsilon^2 \tilde u_{xx} + \tilde f(\tilde u, x, \epsilon t)+
\epsilon g_1(u, u_x, x, \epsilon t),
\end{equation}
where
\begin{equation}
\begin{split}
& g_1(x,t,\epsilon)  =\beta^{-1}(\beta_{\tau}\tilde u + \epsilon
\beta_{xx} \tilde u +2\epsilon \beta_{x} \tilde u_x + \epsilon
{w_0}_{xx} -
{w_0}_{\tau}) \\
&  \tilde f ( \tilde u, x, \tau )  = f (\beta \tilde u + w_0, x,
\tau )
\end{split}
\end{equation}
We notice that $|g_1| < c$, with a constant $c$ uniform in $x, t,
\epsilon$. Below, to simplify formulas, we omit the symbol tilde in
notation.

To prove assertion {\bf I}, we use the comparison principle
\cite{Frid}. Our construction of a supersolution follows from
\cite{FifeMac}, in a modified form, since here, with respect two
\cite{FifeMac}, there are two additional difficulties: smallness of
$\epsilon$ and the nonlinearity $f$ depending on $x, \tau$. Since
$w_1 > w_0$, the function
  $\psi$ defined by Prop.~5.2, is increasing in
$\theta =\epsilon^{-1}(x-q)$ for all fixed $\tau, q$ and thus
$\psi_{\theta}
> 0$.

As a  supersolution $u^+$ we take
\begin{equation}
 u^+=\psi(\theta, q, \tau) + \delta,
\label{fm}
\end{equation}
where $\delta$ is a small number, $\psi$  is defined  by Prop.~5.2
and satisfies
\begin{equation}
-V(q,\tau) \psi_\theta=\psi_{\theta\theta}+ f(\psi, q,\tau).
\label{4. }
\end{equation}

 The function $q$ is defined by
\begin{equation}
  \frac{dq}{dt}=\epsilon (V(q,\tau) -\delta_1),   \quad q(0)=q_0,
\label{evinf1}
\end{equation}
  where  the "unperturbed" speed $V(q,\tau)$ is defined by Prop.~5.2.
Both constants $\delta, \delta_1$  depend on $\epsilon$.

Let us check that $u^+$ satisfies the comparison principle
\cite{Frid}. We must check three
 main inequalities
 \cite{Frid}.
The first, according to Theorem 17 (Chapter II) from \cite{Frid}, at
the boundaries $x=0, 1$ our supersolution must satisfy the
inequalities
\begin{equation}
\beta u^+_x(0,t) + {w_0}_x(0, \tau) + \beta_x(0, \tau) u^+(0,t) < 0,
\label{bound1}
\end{equation}
\begin{equation}
\beta  u^+_x(1,t) + {w_0}_x(1, \tau) + \beta_x(1, \tau)  u^+(1,t) > 0.
\label{bound2}
\end{equation}
  One can check that, under the conditions
$q > c\epsilon^{s_2}$ and
$q < 1 - c\epsilon^{s_2}$,  $s_2 \in (0,1)$,
  both inequalities hold
due to (\ref{condw}) and our explicit formula for $u^+$.   In fact, up to
exponentially small corrections, the left hand side of
 inequality (\ref{bound1}) is $(w_0)_x(0, \tau)$ and the left hand side
of (\ref{bound2}) is $(w_1)_x(1, \tau)$.

The second, at $t=0$ our supersolution must  majorize the initial
data:
$$
       u^+(x,0) > u_0(x).
$$
  This estimate holds  due to  relations (\ref{fm}), (\ref{evinf1})
and (\ref{initcond})
if $\delta = C\epsilon^s$ and $C > \bar c_0$.

The third,
  it is necessary to check the following inequality:
\begin{equation}
0 > (V-\delta_1)\psi_\theta + \psi_{\theta\theta}+ f(\psi+ \delta,
x, \tau)  -\epsilon \psi_q (V-\delta_1) - \epsilon \psi_{\tau} +
\epsilon g_1(\psi+ \delta, \psi_x, x, \epsilon t). \label{4.x}
\end{equation}
Let us notice that
$$
  |\psi_{\tau}|, |\psi_q| < c.
$$
Furthermore,
by the Poincar\'e inequality we have $a \delta_0^2 <
\int_{-\infty}^{\infty} \psi_\theta^2 d \theta,  a>0$ and by
equation (\ref{eqc}), one obtains $|V(q,\tau)| < b / \delta_0^2$.

Using the definition of $\psi$ and the last estimates, let us
replace  inequality (\ref{4.x}) by a stronger inequality
\begin{equation}
0 > -\delta_1 \psi_\theta +
f(\psi+ \delta,x, \tau) -f(\psi, q, \tau) + C_1\epsilon.
\label{4.0}
\end{equation}
This inequality can be rewritten as
\begin{equation}
0 > -\delta_1 \psi_\theta + F_1 + F_2
 + C_1\epsilon,
\label{4.1}
\end{equation}
where
$$
  F_1=f(\psi+ \delta,x, \tau) -f(\psi, x, \tau),
$$
$$
  F_2=f(\psi,x, \tau) -f(\psi, q, \tau).
$$
Let us introduce two domains $\Omega_1$ and $\Omega_2$
depending on $t$.  The domain $\Omega_1$ is an union
of the two intervals
\begin{equation}
  \Omega_1^+=\{q <  x < q-a_1 \epsilon \log \epsilon \},
\end{equation}
\begin{equation}
   \Omega_1^-=\{q + a_0 \epsilon \log \epsilon < x < q \},
\end{equation}
and $\Omega_2$ is a complementary domain
\begin{equation}
   \Omega_2=\Omega -\Omega_1.
\end{equation}
Here the constants $a_i > 0$ are uniform in $\epsilon$ and will be chosen
below together with $\delta, \delta_1$.

We choose these parameters as follows:
$$
  \delta=C\epsilon^s, \quad \delta_1=C_1\epsilon^{s_1}, \quad s, s_1 > 0,
$$
  where $C > \bar c_0$, the numbers $s_1, s$ and $a_i$ satisfy
\begin{equation}
  s_1 < s <  s_1 + a_1 \tilde b_1 < 1,
\label{ss11}
\end{equation}
\begin{equation}
  s <  s_1 + a_0 \tilde b_0 < 1.
\label{ss12}
\end{equation}
It is clear that, given positive $\tilde b_i$,
such a choice of $a_i, s_1, s$ is always possible.
Below
all constants $C, C_k, c, c_k$ are uniform in $\epsilon$ as  $\epsilon \to 0$.

Let us estimate $F_2$. Suppose $x > q$.
One has
$$
  |F_2| \le |f_{\xi}(\psi, \xi, \tau)| |x-q|,
$$
where $\xi \in [q, x]$.
  Since $f(w_1, q, \tau)=0$ for any $q$ and $w_1$ is independent
of $q$ (due to our assumption
in beginning of the demonstration) one has
$f_q(w_1, q, \tau)=0$.


Thus the last estimate gives
$$
  |F_2| \le c \epsilon |\psi(\theta, q, \tau)-w_1||x-q|.
$$
  Using the exponential estimates
$|\psi - w_1| < C\exp(- \tilde b_1 \theta)$ for the interfaces (see
Prop.~5.2) and the fact that $\theta \exp(- \tilde b_1 \theta)$ is a
bounded function on $(0, \infty)$,
 one has finally
$$
  |F_2| \le c_1 \epsilon, \quad   \ x > q.
$$
The same inequality can be derived for $x < q$.
Using this estimate, we  replace  main inequality (\ref{4.1})
to a stronger inequality:
\begin{equation}
0 > -\delta_1 \psi_\theta + F_1
 + C_2\epsilon,
\label{4.1m}
\end{equation}
where  $C_2$ is uniform in $\epsilon$ as $\epsilon \to 0$.

Let us turn into $F_1$. Let $x > q, x \in \Omega_2$.
We use now the following remark (see also \cite{FifeMac}).
If $\rho$ is sufficiently small, for example,
$|\rho| < r_1$, where $r_1$ is independent of $\epsilon$, then
$$
  f_u(w_1 +\rho + \rho_1, x, \tau) < -\mu_1/2
$$
  for any $\rho_1 \in [0, \delta)$. Therefore, one obtains
(due to our conditions on $f_u(w_1, x, \tau)$)
$$
  F_1 < -c_3\delta,   \quad  x \in \Omega_2, \ x > q.
$$

First let us consider (\ref{4.1m}) in the domain $\Omega_2$ for
$x > q$ ( the case $x < q$ can be considered in a similar way).
By replacing $F_1$ by its upper estimate one derives
a stronger inequality
$$
  0 > C_2\epsilon  -   c_3 \delta - \delta_1 \psi_{\theta}.
$$
  This holds due to our choice of the parameter $s$ and
the interface monotonicity  ($\psi_{\theta} > 0$),
  since  for small $\epsilon$ one has
$  \delta=\epsilon^s $   and $\epsilon = o(\delta)$ as
$\epsilon \to 0$.

Let us check  inequality (\ref{4.1m}) in $\Omega_1^+$.

 Thanks to the exponential asymptotics
$$
\psi_{\theta \theta} = - \kappa(q, \tau)\exp(-b_1(q, \tau) \theta) +
O(\exp(-2b_1 \theta)),  \quad \theta \to \infty,
$$
where $\kappa > 0$, the function $\psi$ is convex for sufficiently
large $\theta$, i.e.,
$$
\psi_{\theta \theta} (\theta, q, \tau) < 0, \quad \theta > \theta_0(q, \tau).
$$
Moreover due to (\ref{df}) and the
asymptotics $\psi(+\infty, q, \tau)=w_1$ of the interface $\psi$,
there is a number $\theta_1$ such that
$F_1(\psi, x, \tau) < 0$ for $x > q+ \epsilon \theta_1$.
Let us set $\theta_2=\max\{\theta_0, \theta_1 \}$. Consider the subdomain
$W_+$
of $\Omega_1^+$, where $x > q + \epsilon \theta_2$. In this domain
$F_1 < 0$ and, due to the convexity
$\psi_{\theta \theta} < 0$ in $W$,
the function $\psi_{\theta}$  takes the minimal value
at $x=q - a_1 \epsilon \log \epsilon$.

Let us check now inequality (\ref{4.1m}) for the points $x \in W$.
Due to this property of $\psi$, and our choice of $s, s_1, a_1$,
in $W$ one has
$$
   C_2\epsilon <  c\epsilon^{s_1 + a_1 \tilde b_1} \le
\delta_1 \psi_{\theta}
$$
for small $\epsilon$. Since in this domain $W$ one has $F_1 < 0$,
 main inequality (\ref{4.1m}) holds in $W$.

If $x \in \Omega_1^+$ and $x \notin W$,
then $\psi_{\theta} > \kappa_0$, where $\kappa_0 > 0$
is independent of $\epsilon$.  Therefore,
main inequality (\ref{4.1}) again holds for sufficiently small $\epsilon$.
In fact, $|F_1| < c\delta$ and
$F_1 + C_2 \epsilon=o(\delta_1)$ as $\epsilon \to 0$
under conditions (\ref{ss11}) and (\ref{ss12}) on the choice
$s_1, s$.

 We have proved that
 the function $u^+$
is actually a supersolution. In a similar way, we can construct an analogous
 subsolution.
Thus the theorem is completely proved.





{\bf Remark 2}. There are situations where we need more delicate
methods (than the comparison principle) to investigate the interface
movement. Such situations occur when we study an interaction between
many interfaces (see \cite{Fusco,Carr, MOLV}) or, for the
metastability problem, $V(q) = 0$ for all $q$. The last case
corresponds to $u_0+u_1 \equiv 2 u_2$ in the Allen-Cahn model. For
the classical Allen-Cahn model with constant $A, u_0,u_1,u_2$ and
zero Neumann boundary conditions  it is known \cite{Carr,Pinto} that
interface solutions have exponential life times but that they
unavoidably end up at the boundary. The result {\bf II} does not
hold in this case because $V'(q)=0$. Interfaces can not satisfy the
boundary conditions precisely. They do this only up to terms that
are exponential in the distance between the interface and the
boundary \cite{Pinto}. This heuristic argument explains the
attraction of interfaces by the boundaries. More generally, it is
known that the only stable solution of equation (\ref{eqRD}) with
homogeneous function $f=f(u)$ and with Neumann no flux boundary
conditions is homogeneous \cite{Smoller}.

{\bf Remark 3}. The result {\bf I} can be extended to the case
when there is a number of different interfaces. If $f(u,x,
\tau)=f(u)$,
 the case of many
interfaces was studied in \cite{Carr} and  on a formal level, in
\cite{MOLV}.

\section{Applications to shear banding and morphogenesis}

\subsection{Shear banding in pipe flow}


Shear banding is an instability of complex fluids. It consists in
the development of bands of different viscosities and shear rates
when the fluid is sheared. To study flow of complex  fluids the
Navier-Stokes equations should be coupled to constitutive equations.
In \cite{OlmRadLu,Rad,RadOlmLu} shear banding of worm-like micelles
has been modelled by using the Johnson-Segalman constitutive model,
see also \cite{pego,brunovski,malkus}. Like in \cite{leal} stress
diffusion has been added to cope with transport of order parameter
across inhomogeneous interfaces. In the (singular) limit of small
Reynolds number the dynamics on time scales longer than the total
stress equilibration time is described by a set of two
reaction-diffusion equations. Unfortunately this system is neither
gradient nor monotone, so the results of the preceding sections
cannot be applied to it. In \cite{RadOlmLu} a toy model consisting
of one reaction-diffusion equation has been used to mimic the basic
properties of the phenomenon. Numerical simulations suggest that
this equation has similar properties with respect to the existence
and the propagation of kinks as the system yielded by the
Johnson-Segalman model. In the Poiseuille (pipe flow) flow geometry
the equation reads:

\begin{equation}
S_t = \epsilon^2 S_{xx} - S \left[ 1 + \left( \frac{
\sigma(x,\tau) - S}{\eta} \right)^2 \right] + \frac{
\sigma(x,\tau) - S}{\eta}, \quad \tau = \epsilon t,
\end{equation}
where $S$ is the part of the total shear stress carried by the
micelles, $\epsilon^2$ is the stress diffusion coefficient, $0 <
\eta < 1/8$ is the retardation parameter, i.e., the ratio between
the viscosity of the solvent and the viscosity of the micelles.
$\sigma$ is the total shear stress, which in the Poiseuille geometry
depends linearly on the distance $x$ orthogonal to the pipe axis
$\sigma(x,\tau) = g(\tau) x$, the slow time function $g(\tau)$ is
the pressure gradient that sustains the flow.

The shorted equation for this system is
\begin{equation}S_t = f(S,\eta,\sigma) - S \left[ 1 + \left( \frac{ \sigma - S}{\eta} \right)^2
\right] + \frac{ \sigma - S}{\eta}. \end{equation}

If $0 < \eta <1/8$, the third order polynomial $f(S,\eta,\sigma)$
has one real root which is a stable attractor for (I) $\sigma <
\sigma_1(\eta)$, or (II) $\sigma > \sigma_2(\eta)$, and three real
roots among which two stable attractors $S_0(\sigma),S_1(\sigma)$
and a saddle $S_2(\sigma)$ for (III) $\sigma_1(\eta) < \sigma <
\sigma_2(\eta)$. One can notice that in the case (III) the model is
equivalent to the Allen-Cahn model with the parameters
$A=\frac{1}{\eta}$, $u_i=S_i(\sigma), i=1,3$.

Starting from the pipe axis $\sigma$ increases with $x$ and one
passes from the situation (I) close to the  axis to the situation
(III) and eventually, if the pressure gradient is large enough to
the situation (II) close to the pipe walls. Hence if $g  >
\sigma_1(\eta)/L$ where $L$ is the half width of the pipe, an
interface can form parallel to the pipe axis, separating bands of
high (close to the axis) and low viscosity (close to the walls). The
 interface position $q$ propagates according to the
 following (approximate) equation:

\begin{equation}\frac{d q}{dt} = \sqrt{\frac{2 D}{\eta}}
[S_0(g(\tau) q)+S_1(g(\tau) q) -
 2 S_2(g (\tau) q)].
\end{equation}

If $g$ is not time dependent the interface will be at rest in the
position $q_*$ satisfying $S_0(g q_*)+S_1(g q_*) - 2 S_2(g q_*) =
0$. If such a position does not exist, steady flow will not be
banded. If such a rest point exists, then the asymptotic relaxation
of the interface position is exponential $q - q_* = C_1 \exp
(-t/\tau)$, with $\tau^{-1} = \sqrt{\frac{2D}{\eta}} g [S'_0(g
q_*)+S'_1(g q_*) - 2 S'_2(g q_*)]$. This relation has been used in
\cite{Rad} to estimate $D$ from rheological measurements.

\subsection{Segmentation of the fruit fly}

Let us come back to the examples analyzed in the section 4 and let
us discuss the situations involving diffusion dependent patterning.
The remark from the previous section applies here as well: excepting
the case of monotonous systems $K_{i,j} \geq 0, i\neq j$ there are
no rigorous results in the case of several components ($n >1$).
One could use matched asymptotic techniques in order to obtain
interface velocity estimates for $n>1$ in the limit $\alpha \to
\infty$, but this is cumbersome and will not be discussed here. To
illustrate our method and concepts we shall only refer to the case
of one component ($n=1$).

{\bf Example}

In the gene circuit model with  $n=1$, bistable interface and
diffusion dependent patterning are possible if $K_{11}
> 0$ (gene $1$ activates itself). In this case the interiors of the
intervals  $I^{(1)}$,$I^{(2)}$ overlap. Let $I^{(3)} = I^{(1)} \cap
I^{(2)} = \{ x | \tilde{h}_1 - \tilde{K}_{11} <  m(x)  < \tilde{h}_1
\} $. On the interval $I^{(3)}$ the attractor $S^{(1)}=(1)$ coding
for $u_- = \Ord{\alpha^{-1}}$ and the attractor $S^{(1)}=(1)$ coding
for $u_+ = \lambda^{-1} + \Ord{\alpha^{-1}}$ coexist. An interface
of width of order $\Ord{\epsilon}$ (where $\epsilon=\sqrt{D}$, $D$
being the diffusion coefficient) exists at a position $q \in
I^{(3)}$. Strictly speaking we have proven Theorem 5.2 only in the
case when $I^{(3)}$ extends across the entire domain of $x$.
 This avoids dealing with the extremities of $I^{(3)}$ which are
turning points. If this extra condition is satisfied, we can apply
Theorem 5.2 and equation (\ref{5.4}) giving the kink velocity to
obtain the following approximate equation for the position of the
interface:
\begin{equation}
\frac{dq}{dt} = 2 \lambda \epsilon  \sqrt{2 \lambda } [
\frac{1}{2\lambda} - h + m(q) ]. \label{eqmvt}
\end{equation}
The interface is at rest at a position $q_*$ such that:

\begin{equation}
m(q_*) = h - \frac{1}{2\lambda}. \label{position}
\end{equation}

We should warn the biologist reader that this result is valid
asymptotically in the small diffusion limit $\epsilon \to 0$. It
implies that reaction and diffusion  time scales are well separated.
In this case only, first a kink forms in a position depending on
initial data and then this moves slowly according to the equation
\ref{eqmvt} (see also \cite{FifeHsi}). If the two time scales are
not separated (large diffusion coefficient) then the kink forms and
finds its position in the same time. For large diffusion, the
equilibrium position can be shifted from the position given by
Eq.~\ref{position}.


\subsection{Biological and physical consequences}

Our results on diffusion dependent patterning do not conflict
Wolpert's concept of positional information. Actually, they add
quantitative precision to it. Theorem 5.3 (see equations (5.27,
\ref{position})) backs up Wolpert's theory of pattern formation: the
equilibrium positions of the interfaces between segments are
functions of the morphogens levels. The difference between our
results and Wolpert's theory lies at the level of dynamical details,
most particularly in the diffusion dependent case.

Diffusion neutral patterning does not need diffusion to function. If
the unique attractor condition (3.3) is satisfied, then the pattern
is independent on initial data. Wolpert's threshold mechanism for
positioning interfaces explains the essential features of
patterning.

The major difference between our results and Wolpert's theory occurs
in the diffusion dependent case. In this case, there are two new
features: a) patterning is dependent on initial data and in certain
cases this dependence is discontinuous (the Cauchy problem is ill
posed); b) diffusion  partially or totally removes the ill posedness
of the Cauchy problem. Let us explain in more details these new
features.

As discussed in the beginning of Section 5, without diffusion,
interfaces form in positions where initial data are saddles or
repellers of the shorted equation. For instance, the solutions of
the Allen-Cahn model with small initial data and no diffusion may
contain an arbitrary number of bands and of interfaces. Diffusion
partially lifts this degeneracy. According to Theorem 5.3,
stationary interfaces choose well defined positions. Furthermore,
under the uniform dissipativity condition 3.2, the solutions  of the
full system depend continuously on initial data on bounded time
intervals. Nevertheless, at short times, the distance between the
solution of the full system and the solution of the shorted system
can be small (excepting for neighborhoods of the interfaces by
Theorem 5.2). In practical situations when the asymptotic dynamics
can not be observed or has no biological meaning, the regulatory
effect (the new feature b)) of diffusion can be hard to prove.

The sequence of bands and interfaces is robust in early
\cite{Levine} and also in later stages patterning \cite{vonDassow}
of Drosophila. This observation does not exclude diffusion dependent
mechanisms. Indeed, even in the diffusion dependent case, early and
asymptotic sequence of attractors can be identical if initial data
is judiciously chosen, i.e. by ensuring the right amounts of
maternal proteins. The allowable error for this choice depends on
the size of the attraction basins. Diffusion dependent reasonings
could explain experimental findings \cite{Kosman} showing that the
level and timing of gene expression as well as protein diffusion
have consequences on patterning.




Another refinement of Wolpert's theory  concerns the structure of
the interfaces. Transition layers have widths that depend on the
steepness of the sigmoidal function describing gene interaction.
Biologists quantify this steepness by the so-called Hill coefficient
\cite{Bahram}. Very steep interfaces between patterning segments
were observed at the end of the segmentation process of Drosophila
\cite{Bahram}. Nonetheless, it is difficult to explain an unbounded
increase of the Hill coefficient. On the contrary, widths of
diffusion layers do not depend on the steepness of the interactions
but on the diffusion coefficients. Diffusion coefficients and
diffusion layers widths can be decreased drastically by crowding
effects.

In phase transitions of complex fluids, the dynamical
characteristics of diffusion dependent patterning were proven
experimentally \cite{Rad}. These include multiscale dynamics,
metastability, and reproducible selection (independent on initial
data) of the steady state.




\begin{thebibliography}{vDMMO00}

\bibitem[AC79]{Allen}
S.~Allen and J.~Cahn.
\newblock A microscopic theory for antiphase boundary motion and its
  application to domain coarsening.
\newblock {\em Acta Metall.}, 27:1084--1095, 1979.

\bibitem[BS94]{brunovski}
P.~Brunovsky and D.~Sevcovic.
\newblock Explanation of spurt for a non-newtonian fluid by diffusion term.
\newblock {\em Quart.J.Appl.Math.}, 3:401--426, 1994.

\bibitem[CH58]{Cahn-Hilliard}
J.~W. Cahn and J.~E. Hilliard.
\newblock Free energy of a nonuniform system. i. interfacial free energy.
\newblock {\em J. Chem. Phys.}, 28:258--267, 1958.

\bibitem[CP89]{Carr}
J.~Carr and R.L. Pego.
\newblock Metastable patterns in solutions of $u_t=\epsilon^2 u_{xx} + f(u)$.
\newblock {\em Comm. Pure Appl. Math.}, 42:523 --576, 1989.

\bibitem[DK70]{DalKr}
Yu.~L. Dalezkii and M.~G. Krein.
\newblock {\em Stability of solutions of differential equations in Banach
  spaces}.
\newblock Nauka, Moscow, 1970.

\bibitem[EKL89]{leal}
A.W. El-Kareh and L.G.Leal.
\newblock Existence of solutions for all deborah numbers for a non-newtonian
  model modified to include diffusion.
\newblock {\em J.Non-Newtonian Fluid Mechanics}, 33:257--287, 1989.

\bibitem[FH88]{FifeHsi}
P.C. Fife and L.~Hsiao.
\newblock The generation and propagation of internal layers.
\newblock {\em Nonlinear Anal.}, 12:19--41, 1988.

\bibitem[Fif89]{Fife}
P.~C. Fife.
\newblock Diffusive waves in inhomogeneous media.
\newblock {\em Proc. Edinburg Math. Soc.}, 32:291--315, 1989.

\bibitem[FM77]{FifeMac}
P.~C. Fife and J.~B. McLeod.
\newblock The approach of solutions of nonlinear diffusion equations to
  traveling front solutions.
\newblock {\em Arch. Rat. Mech. Anal.}, 65:335--361, 1977.

\bibitem[FN93]{Fun}
K.~Funahashi and Y.~Nakamura.
\newblock Approximation of dynamical systems by continuous time recurrent
  neural networks.
\newblock {\em Neural Networks}, 6:801--806, 1993.

\bibitem[Fri64]{Frid}
A.~Friedman.
\newblock {\em Partial Differential Equations of Parabolic type}.
\newblock Prentice - Hall, Inc. Englewood Cliffs, N. J., 1964.

\bibitem[Fus90]{Fusco}
G.~Fusco.
\newblock A genetic approach to the analysis of $u_t=\epsilon^2 u_{xx} + f(u)$
  for small $\epsilon$.
\newblock In {\em Problems involving change of type, Lecture Notes in Physics,
  359}. Springer, Berlin, 1990.

\bibitem[GM77]{Gurtin}
M.E. Gurtin and R.C. MacCamy.
\newblock On the diffusion of biological populations.
\newblock {\em Math. Biosc.}, 33:35--49, 1977.

\bibitem[Hen81]{Henry}
D.~Henry.
\newblock {\em Geometric Theory of Semiliniar Parabolic Equations}.
\newblock Springer, New York, 1981.

\bibitem[Hir76]{HirschI}
M.W. Hirsch.
\newblock {\em Differential Topology}.
\newblock Springer Verlag, New- York, Heidleberg, Berlin, 1976.

\bibitem[Hir82]{hirsch83}
M.W. Hirsch.
\newblock Systems of differential equations which are competitive or
  cooperative. i: limit sets.
\newblock {\em SIAM J.Appl.Math.}, 13:167--179, 1982.

\bibitem[Hir90]{hirsch90}
M.W. Hirsch.
\newblock Systems of differential equations that are competitive or
  cooperative. iv: Structural stability in three dimensional systems.
\newblock {\em SIAM J.Math.Anal}, 21:1225--1234, 1990.

\bibitem[Hun99]{Hun}
A.~Hunding.
\newblock Turing structures of the second kind.
\newblock In M.A.J. Chaplain, G.D. Singh, and J.C. McLachlan, editors, {\em On
  Growth and Form}. John Wiley \& Sons, 1999.

\bibitem[HWL02]{Bahram}
B.~Houchmanzadeh, E.~Wieschaus, and S.~Leibler.
\newblock Establishment of developmenal precision and proportions in the early
  drosophila embryo.
\newblock {\em Nature}, 415:798--802, 2002.

\bibitem[KL91]{Levine}
R.~Kraut and M.~Levine.
\newblock Mutually repressive interactions between the gap genes ginat and
  krüppel define middle body regions of the drosophila embryo.
\newblock {\em Development}, 111:611--621, 1991.

\bibitem[Kru66]{Kruzkov}
S.N. Kruzkov.
\newblock Apriori estimate for the derivative of a solution to a parabolic
  equation and some of its applications.
\newblock {\em Sov.Math.Dokl.}, 7:1215--1218, 1966.

\bibitem[KS97]{Kosman}
D.~Kosman and S.~Small.
\newblock Concentration-dependent patterning by an ectopic expression domain of
  the drosophila gap gene knirps.
\newblock {\em Development}, 124:1343--1354, 1997.

\bibitem[Kur84]{Kuramoto}
Y.~Kuramoto.
\newblock {\em Chemical oscillations, waves and turbulence. Springer series in
  Synergetics 19.}
\newblock Springer, Berlin, 1984.

\bibitem[Mei82]{Meinhardt}
H.~Meinhardt.
\newblock {\em Models of biological pattern formation}.
\newblock Academic Press, London, 1982.

\bibitem[MNB91]{malkus}
D.S. Malkus, J.S. Nohel, and B.J.Plohr.
\newblock Analysis of new phenomena in shear flow of non-newtonian fluids.
\newblock {\em SIAM J.Appl.Math.}, 51:899--929, 1991.

\bibitem[MSR91]{Mjol}
E.~Mjolness, D.~H. Sharp, and J.~Reinitz.
\newblock A connectionist model of development.
\newblock {\em J. Theor. Biol.}, 152:429--453, 1991.

\bibitem[Mur93]{Murray}
J.D. Murray.
\newblock {\em Mathematical Biology}.
\newblock Springer, New York, 1993.

\bibitem[MV88]{MOLV}
I.A. Molotkov and S.~Vakulenko.
\newblock {\em Sosredotochennye nelineinye volny (Russian) [Localized nonlinear
  waves]}.
\newblock Leningrad. Univ., Leningrad, 1988.

\bibitem[NPT90]{pego}
J.S. Nohel, R.L. Pego, and A.E. Tzavaras.
\newblock Stability of discontinuous steady states in shearing motions of a
  non-newtonian fluid.
\newblock {\em Proc.Roy.Soc.Edin.}, A 115:39--59, 1990.

\bibitem[ORL00]{OlmRadLu}
P.D. Olmsted, O.~Radulescu, and C.Y.D. Lu.
\newblock The johnson-segalman model with a diffusion term: a mechanism for
  stress selection.
\newblock {\em J. Rheology}, 44:257--275, 2000.

\bibitem[Pin01]{Pinto}
J.T. Pinto.
\newblock Slow motion manifolds far from the attractor in multistable
  reaction-diffusion equations.
\newblock {\em J. Diff. Eq.}, 174:101--132, 2001.

\bibitem[Pom86]{Pom}
Y.~Pommeau.
\newblock Front motion metastability and subcritical bifurcations in
  hydrodynamics.
\newblock {\em Physica}, 32 D:3--11, 1986.

\bibitem[RO00]{RadOlm}
O.~Radulescu and P.D. Olmsted.
\newblock Matched asymptotic solutions for the steady banded flow of the
  johnson-segalman model in various geometries.
\newblock {\em J. non-Newtonian Fluid Mech.}, 91:141 -- 162, 2000.

\bibitem[ROD03]{Rad}
O.~Radulescu, P.D. Olmsted, J.P. Decruppe, S.~Lerouge, J.F. Berret,
and
  G.~Porte.
\newblock Time scales in shear banding of wormlike micelles.
\newblock {\em Europhys. Lett.}, 62:230--236, 2003.

\bibitem[ROL99]{RadOlmLu}
O.~Radulescu, P.D. Olmsted, and C.Y.D. Lu.
\newblock Shear banding in reaction-diffusion models.
\newblock {\em Rheol. Acta}, 38:606 -- 613, 1999.

\bibitem[RS95]{Rein1}
J.~Reinitz and D.~H. Sharp.
\newblock Mechanism of formation of eve stripes.
\newblock {\em Mechanisms of Development}, 49:133--158, 1995.

\bibitem[Rue89]{Ru}
D.~Ruelle.
\newblock {\em Elements of differentiable dynamics and bifurcation theory}.
\newblock Acad. Press, Boston, 1989.

\bibitem[Sma76]{smale-mathbio}
S.~Smale.
\newblock On the differential equations of species in competition.
\newblock {\em J.Math.Biol.}, 3:5--7, 1976.

\bibitem[Sma80]{Sm}
S.~Smale.
\newblock {\em Mathematics of Time}.
\newblock Springer, New York, 1980.

\bibitem[Smo83]{Smoller}
J.~Smoller.
\newblock {\em Shock Waves and reaction-diffusion systems}.
\newblock Springer, New York, 1983.

\bibitem[Sou03]{soule}
Christophe Soul\'e.
\newblock Graphic requirements for multistationarity.
\newblock {\em Complexus}, 1:123--133, 2003.

\bibitem[ST91]{ST}
H.~L. Smith and H.~R. Thieme.
\newblock Convergence for strongly order preserving semiflows.
\newblock {\em SIAM J. Math. Anal.}, 22:1081--1101, 1991.

\bibitem[Tho68]{thom-waddington}
R.~Thom.
\newblock Une th\'eorie dynamique de la morphog\'en\`ese.
\newblock In C.H. Waddington, editor, {\em Towards a theoretical biology
  1.Prolegomena}. Edinburgh University Press, Edinburgh, 1968.

\bibitem[Tur52]{Turing}
A.~M. Turing.
\newblock The chemical basis of morphogenesis.
\newblock {\em Phil. Trans. Roy. Soc. B}, 237:37--72, 1952.

\bibitem[Vak97]{Vak2}
S.~Vakulenko.
\newblock Reaction-diffusion systems with prescribed large time behaviour.
\newblock {\em Ann. Inst. H. Poincar\'e}, 66:373--410, 1997.

\bibitem[Vak00]{Vak5}
S.~Vakulenko.
\newblock Dissipative systems generating any structurally stable chaos.
\newblock {\em Adv. Differential Equations}, 5:1139--1178, 2000.

\bibitem[Vak02]{Vak6}
S.~Vakulenko.
\newblock Complexit\'e dynamique de reseaux de hopfield.
\newblock {\em C. R. Acad. Sci. Paris S\'er. I Math.}, t.335:639642, 2002.

\bibitem[vDMMO00]{vonDassow}
G.~von Dassow, E.~Meir, E.~M. Munro, and G.~M. Odell.
\newblock The segment polarity network is a robust developmental module.
\newblock {\em Nature}, 406:188--192, 2000.

\bibitem[VG03a]{Gen8}
S.~Vakulenko and S.~Genieys.
\newblock Pattern programming by genetic networks.
\newblock In A.~Abramian, S.~Vakulenko, and V.~Volpert, editors, {\em Patterns
  and Waves, Collection of papers}, page 346. S. Petersburg, 2003.

\bibitem[VG03b]{VakGrCR}
S.~Vakulenko and D.~Grigoriev.
\newblock Complexity of gene circuits, pfaffian functions and the morphogenesis
  problem.
\newblock {\em C. R. Acad. Sci. Paris S\'er. I Math.}, 337:721--724, 2003.

\bibitem[VV96]{VVA2}
V.~Volpert and A.~Volpert.
\newblock Wave trains described by monotone parabolic systems.
\newblock {\em Nonlinear World}, 3:159--181, 1996.

\bibitem[VV99]{VVA3}
V.~Volpert and A.~Volpert.
\newblock Existence of multidimensional travelling waves in the bistable case.
\newblock {\em C. R. Acad. Sci. Patis S\'erie I}, t.328:245--250, 1999.

\bibitem[VV01]{VakVol2}
S.~Vakulenko and V.~Volpert.
\newblock Generalized travelling waves for perturbed monotone
  reaction-diffusion systems.
\newblock {\em Nonlinear Anal. Ser. A: Theory Methods}, 46:757--776, 2001.

\bibitem[VVV94]{VVA1}
A.~I. Volpert, V.~Volpert, and A.~Volpert.
\newblock {\em Traveling waves solutions of parabolic systems, Transl. of Math.
  Monographs 140}.
\newblock AMS, Providence RI, 1994.

\bibitem[WBJ02]{Wolpert}
L.~Wolpert, R.~Beddington, T.~Jessell, P.~Lawrence, E.~Meyerowitz,
and
  J.~Smith.
\newblock {\em Principles of Development}.
\newblock Oxford University Press, Oxford, 2002.

\bibitem[Wol70]{wolpert-waddington}
L.~Wolpert.
\newblock Positional information and pattern formation.
\newblock In C.H. Waddington, editor, {\em Towards a theoretical biology
  3.Drafts}. Aldine Publishing Company, Chicago, 1970.

\end{thebibliography}
\end{document}